# The evolving system of Trypillian settlements

Anvar Shukurov [a,1], Mykhailo Videiko [b]

[a] *School of Mathematics and Statistics, Newcastle University, Newcastle upon Tyne, NE1 7RU, UK*
[b] *Institute of Archaeology, National Academy of Sciences, 12 Geroiv Stalingrada, Kyiv, 02410, Ukraine*

## Abstract

Archaeological settlement systems are usually analysed in terms of the relation between the rank of a settlement and its size (Zipf's law). We argue that this approach is unreliable, and can be misleading, in application to archaeological data where the recovery rate of settlements can be low and their size estimates are often approximate at best. An alternative framework for the settlement data interpretation, better connected with the theoretical concepts of the stochastic evolution of settlement systems, is applied to the evolving system of settlements of the Late Neolithic–Bronze Age Trypillia cultural complex (5,400–2,800 BC) in modern Ukraine. The stochastic evolution model provides a consistent and accurate explanation of the frequency of occurrence of Trypillian settlement areas in the range from 0.05 to 500 ha. Thus validated, the model leads to reliable estimates of the typical size of a newly formed settlement as well as the growth rates of the total number of settlements and their areas. The parameters of the settlement system thus revealed are consistent with palaeoeconomy reconstructions for the Trypillia area.

## 1. Introduction

Analysis of archaeological settlement systems can provide unique insight into the population dynamics, exchange networks, palaeoeconomy and social structure of prehistoric societies (Johnson, 1977; Duffy, 2015; see also Pumain and Guerois, 2004). Studies of this kind require systematic data collection from archaeological surveys, excavations, geophysical prospecting, aerial imaging, etc. Among such systems explored earlier are the Uruk settlement system in the Susiana Plain of South-Western Iran (Johnson, 1977) and Basin of Mexico in pre-Hispanic America (Ortman et al., 2015, and references therein). The often adopted framework of these studies is Zipf's law that relates the size of a settlement to its rank in the system, and analysis of plausible exchange and administration networks (e.g., Knappett et al., 2008).

More than 4,500 settlements, burial sites and burials are known for the cultural complex of Cucuteni–Trypillia, with about half of them located in the modern Ukraine. This is one of the best-explored prehistoric phenomenon in Europe, and its studies have been developing fast in the last decade (Diachenko, 2008; Brigand and Weller, 2013). Here we analyse the settlements of the Trypillian culture in the Ukraine.

We use the term 'settlement system' simply to describe a collection of settlements rather than as a reference to any system of rules that generates the site distribution in size or their spatial pattern (cf. Flannery, 1976; Duffy, 2015). This meaning is close to the notions of site size hierarchy or settlement pattern.

---

[1] Corresponding author.
*E-mail address:* <anvar.shukurov@ncl.ac.uk> (A. Shukurov).

The plan of the paper is as follows. After a brief review of the Trypillian cultural complex in Section 2 and its settlements in Section 3, we motivate our approach to settlement systems and compare it with those used earlier in Section 4. Models of the evolution of settlement systems are briefly reviewed in Section 4, where we attempt at explaining mathematical models in an intuitively clear manner to introduce basic concepts, assumptions and implications without relying on the reader's mathematical background. In Section 4.4, we discuss how these models can be applied in archaeology and argue that the earlier analyses of prehistoric settlement patterns are questionable. The application to Trypillian settlements is discussed in Sections 5 and 6, first to the cumulative pattern of the settlements of all stages and then to the individual typo-chronological stages. The interpretation of the results in terms of the evolution of the settlement system (the rate of its expansion or contraction and the typical size of newly formed settlements) can be found in Section 7. Our results are summarized and discussed in Section 8.

For the toponyms in the Ukraine, we use, wherever possible, the spellings of the *Internet Encyclopedia of Ukraine* of the Canadian Institute of Ukrainian Studies (http://www.encyclopediaofukraine.com/) and annotated Google satellite maps (http://www.maplandia.com/ukraine/). The transliteration used are based on the place names in the local languages, Ukrainian and Romanian.

## 2. The Cucuteni–Trypillia culture

The time span of the Trypillian culture extends from the Eneolithic to the Early Bronze Age, about 5400/5300–2900/2750 BC. It represents the eastern component of the cultural complex of Cucuteni–Trypillia that extends across modern Moldova, Romania and Ukraine. It is variously known as Cucuteni–Tripolie/Trypillia, Trypillia–Cucuteni, the Trypillia/Tripolie Culture and the Painted Pottery Culture[2]. It was first identified in the nineteenth century by the Ukrainian archaeologists Khvoika (1904). The Trypillian culture emerged from such Neolithic archaeological cultures as Boian, the Linearbandkeramik, with contributions from Dudeşti, Criş and Hamangia. The influence of these cultures can be traced in the finds from Trypillian settlement of Bernashivka on the Dniester, the earliest in the Ukraine (Zbenovich, 1989).

In modern Ukraine, Trypillian sites have been discovered in 15 administrative provinces ("oblast", out of the total of 24), with occasional finds in four more provinces. They are located within the steppe-forest belt stretching from Transcarpathia in the west to the Dnieper valley in the east. Sites belonging to the later stages also occur in the steppes of the north-western Black Sea littoral. Among the known sites are settlements, flint extraction and processing sites, isolated, in-ground and barrow burials, as well as hoards and isolated finds. Altogether, there are about 2300 such sites in the Ukraine.

Remains of buildings now representing a mass of fired clay are typical of the Trypillian settlements. The pottery can be identified as used for cooking and tableware. The latter often has impressed decorations, encrusted with white or red paste, painted with mineral paints (including monochrome, bichrome and polychrome varieties). Both anthropomorphic (often quite realistic) and zoomorphic sculpture is typical of the culture. A peculiar feature of the Trypillian pottery are models of buildings, sledges, chairs/thrones, battle axes, etc. Implements were made of flint, stone, bone and antler. The Trypillians knew metal industry as evidenced by numerous finds of copper ornaments, implements and weapons. Fabrics were woven on vertical looms. The economy of Trypillia was based on cereal and livestock farming (for a recent review and analysis, see Shukurov et al., 2015). During the later stages, some population groups of the steppe zone turned to nomadic herding (the Usatove culture) (Burdo and Videiko, 2005).

---

[2] 'Trypillia' is a transilteration of the Ukrainian form, whereas Tripolie, Tripolye, Tripolje and Tripol'e are various transliterations of the Russian prononcuation. Painted Pottery Culture (*cultura ceramiki malowanej*) is the term used to describe the Trypillia Culture in Poland in the first half of the twentieth century.

The Trypillian culture had more than 60 identifiable local and chronological groups and varieties. Their names were changing as the chronology was being refined and extended. The earliest Stage A appears to be homogeneous but the later Stage B and C can be finely subdivided. There is also some diversity in the understanding of the interrelationships between various groups and the attribution of sites to a specific group or the Trypillian culture in general. For example, Tsvek (2005) suggested that the settlements in the Southern Bug and Dnieper interfluve represent a separate South-Trypillian culture. Some scholars isolate burials of the Usatovo type into the Usatovo culture considered to be distinct from the Trypillia (Dergachev, 1980).

Table 1. Chronology and areas of Trypillian settlements. All entries refer exclusively to permanent settlements, except for the total number of known Trypillian sites of any size and nature in the third line.

| Typo-chronological stage | A | BI | BI-BII | BII | BII-CI | CI | CII |
|---|---|---|---|---|---|---|---|
| Median date, years BC | 5100 | 4550 | 4200 | 3950 | 3650 | 3400 | 3050 |
| Duration, years | 600 | 500 | 200 | 300 | 300 | 200 | 500 |
| Total number of sites | 53 | 42 | 78 | 208 | 121 | 309 | 253 |
| Number of settlements | 50 | 42 | 74 | 186 | 107 | 250 | 118 |
| Number of settlements with known area | 20 | 15 | 25 | 108 | 46 | 128 | 32 |
| Maximum settlement area, ha | 14 | 60 | 150 | 261 | 150 | 341 | 160 |
| Mean settlement area, ha | 3 | 10 | 32 | 165 | 14 | 23 | 15 |
| Median settlement area, ha | 2 | 5 | 7 | 3 | 10 | 9 | 3 |
| Minimum settlement area, ha | 0.03 | 0.03 | 0.30 | 0.01 | 0.50 | 0.25 | 0.02 |
| Total area of the settlements, ha | 61 | 148 | 800 | 1721 | 642 | 2974 | 464 |

## 2.1. *Periodisation and absolute chronology*

The periodisation of the Trypillian culture is based on the ceramics typology, stratigraphy and absolute chronology. According to Passek (1935) and later authors, the main identifiable stages are the earliest, middle and late Trypillia, referred to as Trypillia A, B and C, respectively, each subdivided into several phases as shown in Table 1: Stage B includes BI, BI-II and BII and Stage C is represented by CI and CII. Some authors (e.g., Dergachev 1980 and others) consider only CII to belong to the late stage and attribute CI to the middle stage, also known as the 'developed Trypillia'. In this text, we refer to Trypillia A, B and C as stages, whereas the finer divisions are called phases.

According to the absolute chronology based on $^{14}$C age determinations, the individual stages and phases vary from 200 to 600 years in duration. Several distinct types of archaeological artefacts that existed consecutively can occur within a single phase. A reason for this is that the lifetime of most Trypillian settlements is rather short, probably from 80–100 to 150 years, after which the settlement was burned out and the population relocated (for details, see Burdo, 2003; Burdo et al., 2013). Only after a few hundred years, the old site could be repopulated. Multi-layered archaeological sites usually have occupations layers belonging to different phases and even stages. Only seasonal settlements and stations have occupational layers constrained to a single phase.

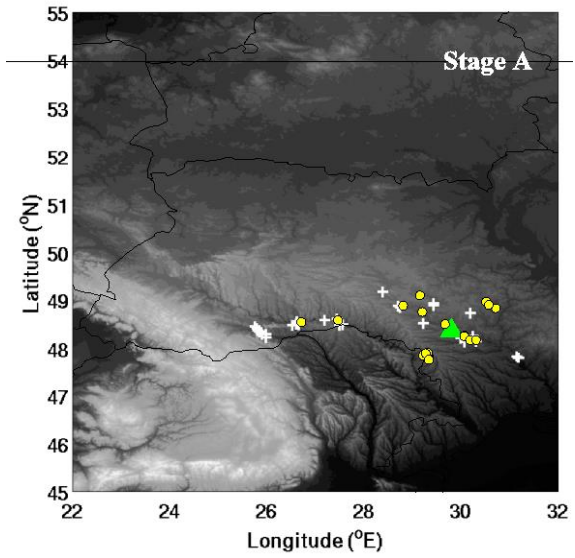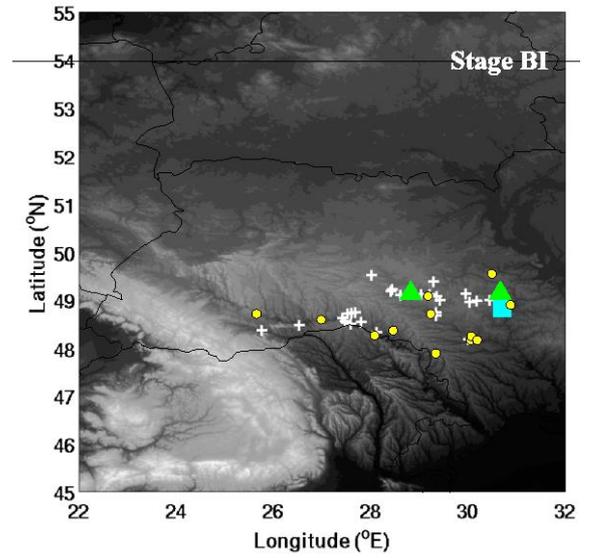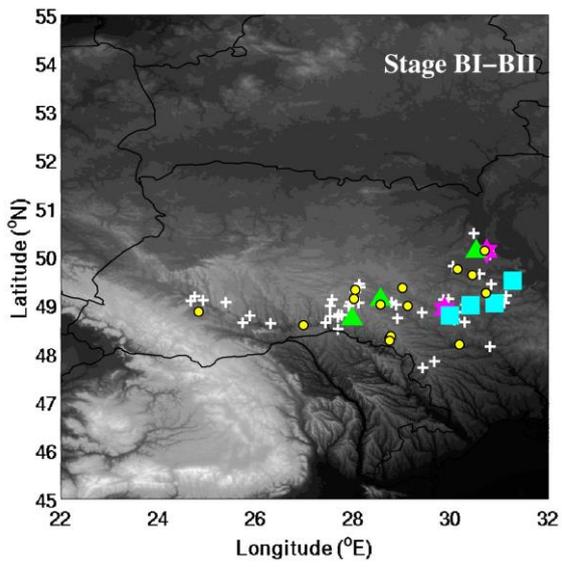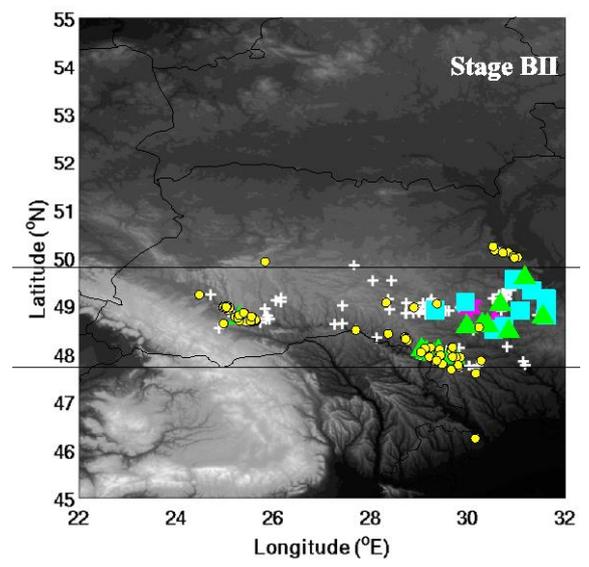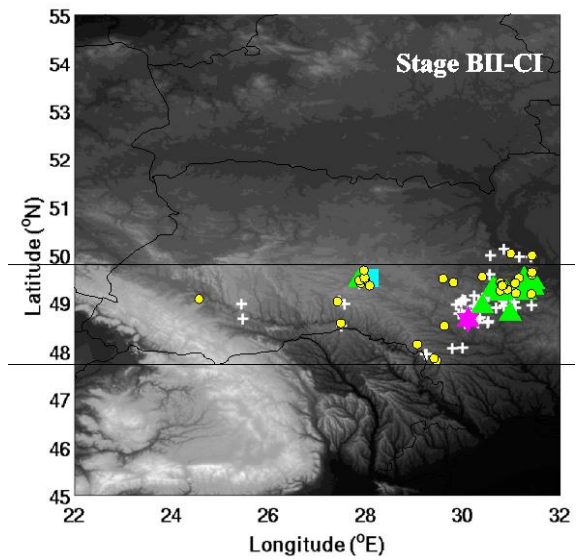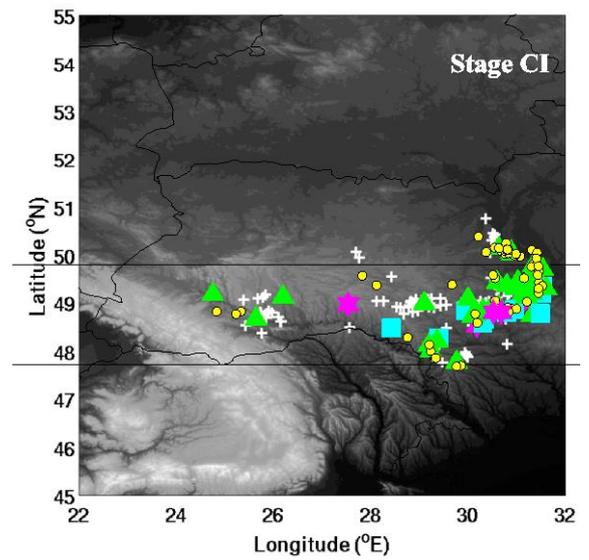

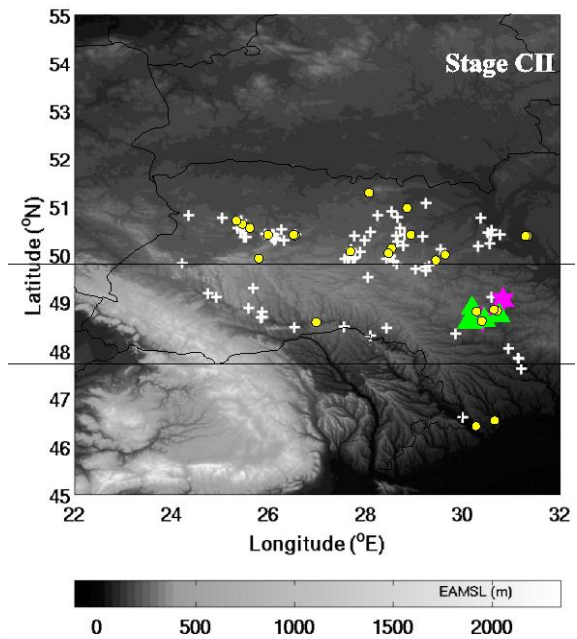

**Figure 1**. The evolution of the system of permanent Trypillia settlements in the Ukraine: Stages A, BI, BI-BII, BII-CI and CII (from left to right and top to bottom). Cross: area unknown; circle: $0 < S \leq 10$ ha; triangle: $10 < S \leq 40$ ha; square: $40 < S \leq 100$ ha; hexagram: $S > 100$ ha. Elevation above mean sea level (EAMSL, in metres) is shown with shades of grey; modern country borders are shown black. The Carpathian Mountains are in the lower left corner; the Black Sea is at the right bottom.

## 2.2. The origin and evolution

The cultural complex of Cucuteni–Trypillia represents an eastern part of the archaeological farming world of Old Europe. The emergence of this complex in the Transcarpathia and in the Dniester basin is dated to the late sixth and early fifth millennia BC and related to such cultures as the Linearbandkeramik (LBK), Boian, Criş and Hamangia and Vinča (Videiko, 2015). These connections are attested by similarities in ceramics, buildings and sculptures. The earliest known settlement in the Ukraine is Bernashivka, located in the middle course of the Dniester. Latest explorations have revealed a few tens of buildings (rather than just seven as suggested by earlier studies) and has several occupation periods, all belonging to Trypillia A. The ceramics assemblage and the style of its decoration suggest that it corresponds to the second phase of Pre-Cucuteni in Romania (Zbenovich, 1980).

Settlements coeval with the third phase of Pre-Cucuteni are spread further up to the right bank of the Southern Bug River where they mostly cluster into groups. Mogil'na III is the largest settlement in the Southern Bug region with the area exceeding 10 ha established with magnetic prospecting. This pattern suggests the dominant west–east direction of the spread.

In the late middle period, Phase BI at about middle of the fifth millennium BC, the Trypillan population spread further east and reached the middle course of the Dnieper River in Phase BI-II. The first to emerge was a group of 15 settlements; the eponymic Trypillia is the largest among them at about 100 ha in area. At the same time, the Bug–Dniester interfluve was the location of not only numerous relatively small settlements but also major centres such as Veslyi Kut (about 150 ha) and Miropolye (about 200 ha). At the turn of the fifth–fourth millennium BC, the Trypillians emerge on the left bank of the Dnieper where they occupied a belt along the river of not more than 10–20 km on each side of the watercourse, with only isolated settlements located outside it. The southern limits of the Trypillian culture are at the Kodyma and Syniukha basins, the latter a tributary of the Southern Bug. It is worth a special note that isolated artefacts produced by the Trypillians (ceramics and figurines) were found far beyond the limits of this area, from the steppes of the Black Sea littoral and Northern Caucasus to regions beyond the Western Bug. However, these finds attest to external links rather that to the occupation area of the Trypillians.

# 3. Trypillian settlements

The fifteen decades of the studies of the Trypillian culture have produced vast amount of information on the character and classification of its settlements. Its understanding evolved continuously (Videiko, 2013). In this section, we briefly review those aspects of this information which relevant to this work.

*3.1.* Permanent and temporary settlements

The cultural complex of Cucuteni–Trypillia is represented by both permanent and temporary settlements. The former were the foci of economic activity, with buildings in use through a whole year, inhabited without interruption for many years. Temporary settlements and stations were used for the traditional activities of hunting and fishing as well as for the novel occupations of seasonal pasture herding, mining, travel along stable routes, etc.

The classification of a settlement as permanent or temporary is based on archaeological evidence, primarily the types and diversity of the buildings as well as the features of the cultural layer in general. The buildings of permanent Trypillian settlements are represented by:

1. Remains of burnt-out buildings made of wood and clay. The latter has been preserved as clumps of various sizes that bear impressions of wooden structures. Among such remains, some parts of the interior are occasionally found, such as open hearths and ovens, various types of paving, and wall-plaster from walls and ceilings. Such sites usually yield large quantities of sherds, from ten to hundred and more (Muller and Videiko, 2016a,b; Burdo et al., 2013).
2. Pits with remains of open hearths filled with ceramic fragments and animal bones.

The remains of the first type are interpreted as aboveground wattle-and-daub buildings plastered with cob. Among such building, those used for habitation, household, cult and production purposes can be identified. They were one and two-storey high, and their construction required significant amounts of materials (primarily clay and wood) and labour. When such a building was being abandoned, a special ritual was apparently performed during which the building was provided with ceramics, food supplies (cereal grain and meat), tools and cult implements, and then burnt. Settlements with buildings of this type are accepted as permanent.

Objects of the second type are understood as pit-houses of various depth. When they were abandoned, the pits were gradually filled up with a cultural layer. In some cases, they are believed to be connected with sacrifice rituals (Burdo, 2006; Muller and Videiko, 2016a). New structures of the same type could be constructed next to the abandoned ones. Settlements with this type of buildings are usually treated as temporary.

On the other hand, although settlements usually consist of either substantial buildings or of pit-houses (Khvoika, 1904), some sites contain both. An early understanding was that the pits that occur in settlements with remains of adobe buildings were the pit-houses that were used at an initial period of the settlement existence. An example of such an object is the 'pit-house' discovered in 1939 at the Trypillian settlement Volodymyrivka. A later careful analysis of the field notes revealed, however, that the character of the pit does not fit the description of a pit-house as it has a funnel-shaped profile (Fig. 466 in Passek, 1949), which makes it not suitable for a dwelling.

Recent explorations involving magnetic prospecting have shown that virtually every aboveground building at a Trypillian settlement has an associated trench, a source of clay used for the construction and repairs of the building. Such trenches have been discovered at many settlements, for example, Nebelivka, Maidanetske, Talianki, Dobrovody and many others (Chapman et al., 2014; Videiko et al., 2014).

There are also settlements consisting exclusively of subterranean and, perhaps, aboveground buildings constructed without the use of clay. The settlement Chapaevka in the middle Dnieper region is an example: not less than five dug-out structures have been excavated, surrounded by a few tens of various pits. The dwellings of this settlement were suggested to represent a dug-out with a lobby, covered by a pyramidal roof.

A combination of permanent and temporary settlements is typical of only that part of the Cucuteni–Trypillia that is located close to and along major rivers, the Dniester, Southern Bug and Dnieper. The temporary settlements apparently served to exploit the resources of a river valley suitable for not only hunting and fishing but also providing pasture. Analysis of the faunal remains from the temporary settlements on the Dnieper demonstrates that 80% or more of the diet was the hunted meat, whereas the synchronous permanent settlements have the opposite relation of the meats of domesticated and wild animals. This fact might also suggest that the temporary settlements were seasonal being occupied at least from spring to autumn when domesticated animal could graze on the meadows (Videiko, 2015).

Another case of seasonal occupation is associated with the sources of flint in the Dniester region and Volyn, most importantly in the valley of the Horyn river. No traces of permanent buildings were discovered at the Bodaky site despite many years of excavation although the remains of burnt-out, structures plastered with clay were found in pits (Skakun, 20005:64–66). Sites with evidence of production and processing of flint in Transcarpathia are represented exclusively by mines and processing workshops where the cultural layers containing mostly flint flakes also bear Trypillian ceramics (Vasilenko, 1989).

It is also notable that aboveground structures built with large amounts of clay disappear in the last quarter of the fourth millennium BC. The tradition of burning out buildings before abandoning them also ceases. The cultural layer at such sites contains various dugouts sometimes interpreted as pit-houses. However, a more detailed analysis of the excavation results indicates that these were sources of clay, later filled with fragments of pottery and animal bones, rather than dwellings. It remains unclear whether such settlements are permanent or temporary.

### *3.2. The areas of an archaeological site and a settlement*

Until the early 1970's, the areas of the Trypillian settlements had been determined using archaeological methods alone, i.e., excavations and exploration of surface finds. In the case of Trypillia, the most important artefacts used for this purpose were fired adobe remaining from buildings and fragments of pottery. The area of the Volodymyrivka settlement was thus determined by exploring the area of surface finds complemented with identification of burnt-out buildings using metal probes.

Aerial photography was first applied for this purpose in the early 1970's by Shishkin (1973, 1985) (see also Shmaglij et al., 1973). Despite some uncertainties in the interpretation of the photographs, it became possible for the first time to delineate major settlements of tens and even hundreds of hectares in area. The results were later confirmed with traditional archaeological methods and magnetic prospecting. A combination of those three methods has become standard and is at use at present (Chapman et al., 2014).

The most precise and reliable is magnetic prospecting. By the mid-1990's, a few dozen Trypillian settlements in the Ukraine and Moldova were mapped, in the area from the Prut river to the middle course of the Dnieper river (Dudkin and Videiko, 2009). These included the whole range of settlements, from small (2–5 ha) to large (from 10 to 340 ha). However, the limitations of the prospecting technology of the time and sparse coverage (measurements taken at the vertices of squares of 4×4 or 3×3 $m^2$ in size), targeted at the remains of burnt-out buildings, prevented the discovery of all objects of a settlement. The settlement area suggested by surface finds could be either smaller or larger than the true area.

Another notable aspect is that a settlement area obtained from magnetic prospecting was calculated by just multiplying the length of the area with the detected signal by its width without any attention to the shape of the area; this practice was critically assessed by Diachenko (2010) and Harper (2012).. Diachenko used the area of an ellipse to suggest an area of 341.5 ha for Talianki. Despite its higher accuracy, the widespread deviations from a perfectly elliptical boundary of a settlement still lead to inaccuracies. For example, the area of Talianki estimated as the area of a perfect ellipse differs from the actual area by tens of hectares. However, there are no obstacles in measuring the areas of mapped settlements using bespoke software (Harper, 2012) or standard GIS facilities. For example, Harper (2012) used the maps of the settlement obtained with a variety of techniques, from magnetic prospecting to satellite images to estimate the area of Talianki as 335 ha with the 3–6 ha accuracy. Interestingly, the largest value of the area was obtained from a satellite image and the smallest, from the published geomagnetic map. It remains, however, unclear, how Harper identified the boundary of the settlement that is now overlapped by a modern settlement in the valley of the Talianka river tributary.

Another aspect of the size of a settlement is the number density of buildings (i.e., the number of buildings per unit area) in the built-up area. There were repeated attempts to estimate the average building number density from the maps of such settlements as Maidanetske and Talianki (Kruts, 1989; Videiko, 1992). However, geomagnetic maps obtained more recently demonstrated that the number of buildings in major settlements does not always correspond in a universal manner to the size of a settlement and their arrangement within the built-up area.

In both Nebelivka (Phase BII) and Maidanetske (CI), buildings are arranged along distorted, nested ellipses with a large (a few tens of hectares) area in the centre free of buildings, well-defined streets and quarters within the built-up area and large communal buildings at special locations. However, the similarity does not extend beyond this. About 1300 buildings have been discovered in the area of 238 ha in Nebelivka (Burdo and Videiko, 2016), whereas their number in Maidanetske can be as large as 2900 in the area of 200 ha (Rassmann et al., 2016). The number of buildings in Talianki, the largest known Trypillian settlement, can be as large. Thus, the number density of buildings in Maidanteske, and in the still larger Talianki, is twice as large as in Nebelivka while the areas of these settlements are comparable. Correspondingly, these settlements could play different roles in the social and economic systems of their periods despite their similarly large areas with populations of thousands.

A question unavoidable in the case of large settlements is whether all the buildings were contemporaneous. Excavations of Trypillian settlements and the types of ceramics recovered, demonstrate that, in both small and large settlements, the buildings existed (and were burnt) practically simultaneously (Muller et al., 2016). On the other hand, some buildings could be constructed and then burnt before the majority of structures. This lent support to earlier suggestions of a discernible, structured evolution of such settlements (Smaglij and Videiko, 1993). These suggestions were recently confirmed by large-scale radiocarbon dating of Maidanetske objects (Muller et al., 2016). It should be noted, however, that extensive series of radiocarbon dates for large Trypillian settlements remains in contradiction with the relative chronology based on the typology of painted pottery (Videiko, 2016: 64–67). It appears that the dating accuracy provided by radiocarbon dates available is still insufficient to address such problems.

*3.3.   The accuracy of the settlement area measurements*

The accuracy of the area measurements for Trypillian settlements remains only modest. Field observations suggest that modern agricultural activities can displace pottery fragments and building remains by 100 and more metres. Moreover, in the case of large settlements of tens and hundreds hectares in area, it is difficult to decide whether the evidence suggests a single settlement or a group of smaller villages. It appears that magnetic prospecting and/or high-quality satellite images can help to resolve the problem. Earlier experience shows that even excavations dot not provide confidence that a settlement has been fully recovered, except for those cases where the area is limited by natural borders such as a river or a gully. Trypillian settlements in the middle Dnieper provide a glowing example. Several excavations seasons at Kolomijschina I near the village Khalep'ya revealed about 40 building which were believed to be the complete settlement (Passek and Krychevskij, 1946: Figs. 1–2). This settlement has become a reference example, and the ideas of the character of Trypillian settlements relied on it for decades. However, magnetic prospecting at nearby settlements in 1992 and, later, on the site of old excavations demonstrated convincingly that settlements of this type are about twice as large and do not have a circular plan as thought earlier (Videiko, 2016).

At present, computerised area determinations, that provide a satisfactory accuracy, are only possible for those settlements that have been mapped using magnetic prospecting and are georeferenced. Such settlements represent just 2% of the total number. The area estimates used in this paper have been obtained by a variety of methods. We note that more advanced methods usually lead to smaller areas than those obtained archaeologically.

We have compiled data for 1064 Trypillia sites, with areas known for 374 permanent settlements shown in Figure 1. Archaeological dating of well-stratified sites, pottery typology, together with $^{14}$C and archaeomagnetic dating indicate that the lifetime of a permanent Trypillian settlement was 50–100 years (Krutz, 1989; Markevich, 1981; Telegin, 1985). Thus, not all settlements attributed to a certain phase of 200–600 years in duration existed simultaneously. These estimates are consistent with the suggestion that depletion of soil fertility within the exploitation area of a settlement was a factor in limiting its lifetime (Shukurov et al., 2015). Early pre-Hispanic villages in the South-Western USA (Pueblo I, 770–890 AD), notable for the high quality of the archaeological data (high precision of the dating in particular), apparently had a similar occupation time of 10 to 70 years (Wilshusen and Potter, 2010).

## 4. Settlement size patterns and their interpretation

Similarly to other cultural entities based on a farming economy, most of the archaeological sites of the Cucuteni–Trypillia cultural complex are the remains of permanent settlements (e.g., Pashkevich and Videiko, 2006; Videiko et al., 2004). They represent almost 80 percent of the total number of Trypillian sites discovered in modern Ukraine, and the areas of about 45 percent of the permanent settlements are known to at least some degree of certainty.

The sizes of modern and historic urban centres and smaller settlements are known to be strikingly regular as they follow Zipf's law (for a review, see Gabaix and Ioannides, 2004) that relates the rank of a settlement to it size,

$$R = kN^{-\alpha}, \quad \text{with } \alpha = 1, \tag{1}$$

where $R$ is the settlement's rank (with the settlements ordered by size, the largest settlement has the rank of one, the next largest one having $R = 2$, etc.), $N$ is its size (usually taken to be the population size), and $k$ is a constant. This remarkable regularity has been observed in a wide range of settlement systems, from modern urban agglomerations (e.g., Rozenfeld et al., 2011; Berry and Okulicz-Kozaryn, 2012 and references therein) to historic and prehistoric settlements (e.g., Johnson, 1977, 1980; Kohler and Varien, 2010: 51–56). Convincing explanations of Zipf's law emerged only rather recently; we review them in Section 4.2. A simple power-law relation between the rank and size is expected to occur only for settlements whose size exceeds a certain threshold and under certain, albeit quite general conditions, most importantly, Gibrat's law discussed in Section 4.1 (e.g., Pumain and Guerois, 2004). Along with some other power-law relationships (Gabaix, 2009; Ortman et al., 2014, 2015), Zipf's law (or its variations) is a manifestation of generic social and economic features of human societies (independent of the specific socio-economic conditions and of the level of technical, social and cultural development), as well as of basic laws of probability. As a result, Zipf's law emerges in a broad range of contexts (Reed, 2001; Reed and Hughes, 2002b) as long as sufficiently large areas and time spans are considered (Batty, 2006).

Deviations from Zipf's law in modern data are relatively weak and often difficult to detect with confidence. When such deviations are identifiable, the rank–size distribution still follows closely a power law in a broad range of sizes, but with the exponent $\alpha$ only slightly different from unity (Gabaix and Ioannides, 2004). Since the mechanisms behind Zipf's law are independent of the specific socio-economic conditions and the level of technical, social and cultural development, there are all reasons to expect a similar regularity to occur in prehistoric settlement systems. The rank–size analysis has been applied to a number of archaeological cultures but, surprisingly, the results most often deviate very significantly from Zipf's law, and the rank–size dependence often appears to be far from a simple power law (e.g., Johnson, 1977, 1980; Savage, 1997; Drennan and Peterson, 2004; Kohler and Varien, 2010; Griffin, 2011). Weak integration within the settlement system has been claimed if the settlements of intermediate ranks are larger than what is predicted from Zipf's law (such a distribution is often called *convex*). Otherwise, a strongly centralized system is suggested if the settlements of intermediate ranks are significantly smaller than what follows from Zipf's law (a *primate* or *concave* rank–size distribution). However, several complications arise in the analysis of archaeological settlements. Most importantly, incomplete recovery of archaeological sites and inaccurate estimates of the settlement areas (often distinct from the archaeological site areas) can invalidate the rank–size analysis that relies on a reliable knowledge of a relatively small number of settlements at the large-size end of the settlement system. We discuss these problems in Section 4.4.

In this paper, we use an alternative approach to the interpretation of a system of prehistoric settlements based on the theory of the stochastic evolution of settlement systems. In particular, we demonstrate that apparent deviations from a power-law dependence in prehistoric contexts can arise from ill-suited *representations* of the data. Deviations from Zipf's law are more likely to arise not from any over- or under-centralization but rather from the dynamics of the settlement system, especially the variations of the rate of its expansion and growth in space and time. As a result, we suggest a different framework for the interpretation of archaeological settlement systems solidly based on the theory of settlement evolution briefly reviewed in Sections 4.1 and 4.2. We apply these ideas to the Trypillian settlements in modern Ukraine.

The theory of settlement evolution has been developed in application to modern urban systems where the city population is a natural and readily available measure of the city size. In archaeological data, population is unknown and the only applicable measure of a settlement size is the area of the archaeological site. We use the site area $S$ for the settlement size and discuss in Section 4.3 its relation to the population size.

*4.1. Gibrat's law*

Consider a settlement of a size (area) $S$ that varies (grows or decreases) in time $t$ in proportion to its size. In other words, the increment in the size over a (short) time interval $dt$, denoted $dS$, is proportional to $S$. A similar growth of the population size occurs if the difference between the birth and death rates (with the immigration and emigration rates included) is constant. Then the area increment $dS$ is proportional to both $S$ and $dt$, that is $dS = \mu S \, dt$, or, more conveniently,

$$\frac{dS}{S} = \mu \, dt, \qquad (2)$$

where $\mu$ is known as the growth rate of a settlement. Positive values of $\mu$ mean growth ($dS > 0$) and negative increments ($dS < 0$ and $\mu < 0$) correspond to a reduction in size. In the simplest case, the size of each settlement grows (or decreases) at the same rate as the population of the region as a whole and then the value of $\mu$ is the same for all the settlements.

The growth rate can often be assumed not to change in time either, if a sufficiently short time span is considered. Then Equation (2) can be solved to show that the size of each settlement evolves exponentially, as $S(t) = S_0 \exp(\mu t)$, where $S_0$ is the size at a time $t = t_0$, the start of the evolution. If $\mu$ varies with $t$, which is a plausible situation, this solution is readily generalized to $S(t) = S_0 \int_{t_0}^{t} \exp[\mu(t')] \, dt'$. Such a growth or decrease is similar to the variation of consumer prices under inflation ($\mu > 0$) or deflation ($\mu < 0$) with $\mu$ the inflation (deflation) rate. With $\mu$ independent of $S$, Equation (2) represents Gibrat's law of *proportionate growth* since the increment in the settlement size $dS = \mu S \, dt$ is proportional to $S$.

It is more realistic to allow for the fact that different settlements of the same size can evolve differently because of migrations, different environments, variations in the population birth and death rates in space and time, etc. Such relatively short-term variations are difficult or impossible, and often unnecessary, to describe in full detail. It is more useful to assume that the growth rate $\mu$ is not a constant but varies at random from one settlement to another and from one period to another. It is natural to expect that the growth rate has some definite mean value $\mu$ and a random part. The latter is modelled as a Gaussian random variable with variance $\sigma^2$ and, in the simplest case, both $\mu$ and $\sigma$ can be assumed to be the same for all the settlements of a given epoch. Then Equation (2) changes into

$$\frac{dS}{S} = \mu \, dt + \sigma \, dw(t),  \qquad (3)$$

where the first term on the right-hand side describes, as above, a systematic growth at the rate $\mu$ while the second term is responsible for the random fluctuations in the growth rate of a magnitude quantified by $\sigma$. The factor $dw(t)$ varies randomly with $t$ having the mean value equal to zero and unit variance. In applications to economy and finance, $\sigma$ is called the volatility (e.g., of share prices). The theory remains applicable if both $\mu$ and $\sigma$ vary with time, but at a rate slower than the random variations represented by $dw(t)$. For example, $dw(t)$ could represent time variations within an archaeological stage whereas the variations between the stages could be accounted for by changes in $\mu$ and $\sigma$. The value of the random increment $dw(t)$ at a time $t$ is assumed to be statistically independent of its earlier (or later) values, can be positive or negative, and the mean value of the random increments vanishes.

Because of the random fluctuations in the instantaneous growth rate, the sizes of the settlements are random quantities, and it is convenient to characterize the settlement system in terms of $p(S)$, the probability density of the settlement size $S$, defined such that the number of settlements of sizes between $S$ and $S + dS$ is equal to $Mp(S)$, where $M$ is the total number of settlements. The quantity $Mp(S)$ is known as the frequency of occurrence of the settlements of areas between $S$ and $S + dS$. Equation (3) can be solved to derive the probability density of $S$ if the other parameters appearing in the equation are known. Over a sufficiently long time span, the random increments $\sigma \, dw(t)$ add up to a Gaussian random variable with zero mean value and the variance $\sigma^2 t$ (the central limit theorem). Since $dS/S = d\ln(S)$, this means that $\ln(S/S_0)$, with a certain initial area $S_0$, is a Gaussian (or normal) random variable with the evolving mean value $\mu t$ and standard deviation $\sigma\sqrt{t}$. In other words, the settlement areas have the lognormal probability distribution. If the growth rate varies with time, the mean settlement area is given by (e.g., Gardiner, 2009: 109)

$$\bar{S}(t) = \bar{S}_0 \exp\left(\int_{t_0}^{t} \mu(t') \, dt'\right), \qquad (4)$$

where $\bar{S}_0$ is the mean size at an initial time $t_0$. The mean-squared deviation of the values of $S$ from the mean area $\bar{S}$, denoted $\sigma_S^2$, increases with time as

$$\sigma_S^2 = \overline{(S^2 - \bar{S}^2)} = \bar{S}_0^2 \exp\left(\int_{t_0}^{t} \sigma^2(t') \, dt'\right); \qquad (5)$$

here and elsewhere, overbar denotes quantities averaged over the fluctuations in the settlements sizes at a time $t$.

Figure 2 illustrates the resulting stochastic evolution of settlement sizes obtained as the solution of Equation (3) for $\mu$ and $\sigma$ independent of $t$, which has the form (e.g., Gardiner, 2009)

$$S(t) = S_0 \exp[(\mu - \sigma^2/2)\Delta t + \sigma N(0,1)\sqrt{\Delta t}], \qquad (6)$$

where $\Delta t = t - t_0$ and N(0,1) is the Gaussian random variable of zero mean and unit variance. Each curve in Figure 2 represents a possible evolution of the area of a separate settlement. The initial area is the same in all cases shown in the figure but it presents three groups of settlements found at different times, 0, 300 and 600 years. As a result of the random fluctuations, the sizes of individual settlements, identical initially, diverge with time from the mean and from each other at a rate depending on $\sigma$. The individual settlement 'trajectories' are bundled into three groups by their founding times only for the purpose of illustration: in reality, the foundation times of the settlements do not need to be so restricted and can also be random. For the sake of illustration, we also do not show terminated trajectories of settlements that die during the evolution. In reality, the lengths of individual trajectories are also random as some settlements cease to exist at random times. As argued above, before equation (4), the probability distribution of the settlement sizes is lognormal at any given time within each group of curves, as follows from Equation (6), but the probability distribution of the whole system of settlements is a mixture of lognormal distributions. As we discuss below, such a mixture is a power law in $S$ if the settlements are found at random times.

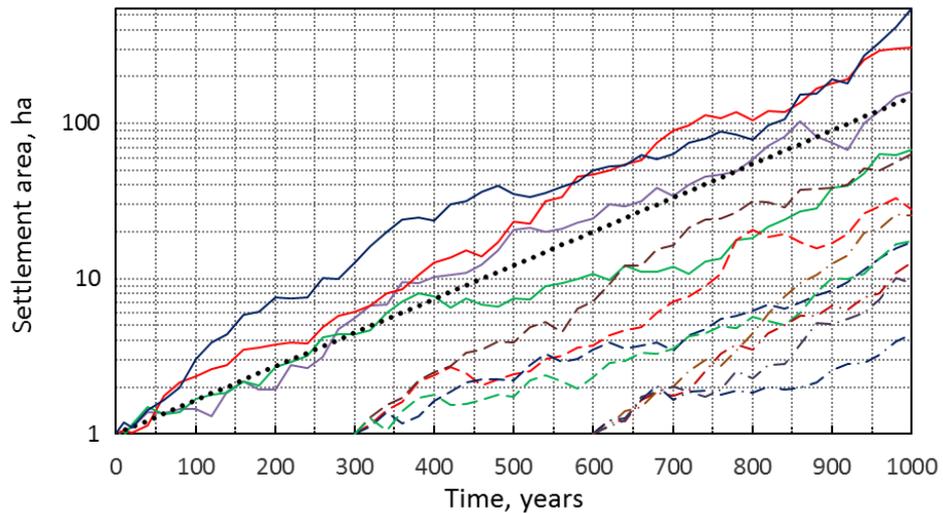

Figure 2. An illustration of the evolution of settlement sizes governed by Gibrat's law. Shown are the areas of settlements varying with time as in the solution of Equation (3) given in Equation (6). The initial area is $S_0 = 1$ ha in all cases shown, but the evolution starts at $t_0 = 0$, 300 and 600 years for the three bundles of curves shown with solid, dashed and dash-dotted lines, respectively. The dotted line shows the mean area $\bar{S} = S_0 \exp[\mu(t - t_0)]$ for the bundle starting at $t_0 = 0$; similar mean dependencies for the other bundles have the same slope and only differ in their starting points. The solutions shown are for the growth rate $\mu = 1/(200 \text{ years})$ and the standard deviation $\sigma^2 = 1/(30,000 \text{ years})$, chosen for illustrative purposes. With the logarithmic scale on the vertical axis and a linear scale on the time axis, an exponential variation is represented by a straight line.

## 4.2. Zipf's law

The regularity in the ranks and sizes of the settlements is only an approximate expression of Zipf's law (or a more general power-law dependence), better suitable for presentation purposes than for quantitative analysis (Gabaix, 1999: Section III.4; Gabaix and Ioannides, 2004: Section 2.1). It is more useful to formulate Zipf's law in terms of the probability for a settlement to have a certain size. The fact that the size $S$ of a settlement is related to its rank $R$ as in Equation (1) (written for $S$ rather than $N$, as in Section 5.1) means that the probability for a settlement to have a size larger than $S$ is given by

$$P(S) = aS^{-\alpha}, \qquad \alpha = 1, \tag{7}$$

where $a$ is a constant. More precisely, $R(S) = M[1 − P(S)] + 1$, where $M$ is the total number of settlements (e.g., Nam and Reilly, 2013). Equation (7) reformulates Zipf's law in terms of the probability $P(S)$, the *cumulative probability distribution* of the settlement sizes. The cumulative probability distribution is equal to the integral of the corresponding probability density: $P(S) = \int_0^S p(S')\,dS'$. The cumulative probability is equal to the fraction of settlements whose size is less than $S$; correspondingly, $MP(S)$ is the number of sites smaller than $S$ in area.

In more convenient to use the probability density, the derivative of the cumulative distribution, $p(S) = dP/dS$. (The probability density can be obtained from a histogram of the site sizes by dividing the frequency in each bin by the total number of sites.) This leads to Zipf's law formulated in terms of the probability density of the settlement size:

$$p(S) = aS^{-(\alpha+1)}, \quad \alpha = 1, \quad a = \alpha k. \tag{8}$$

It is important to keep in mind that Zipf's law is an accurate representation for the sizes of large settlements, that is, Equation (8) approximates the data only for large values of $S$ (or low ranks $R$). Systematic deviations from Zipf's law are often observed at small values of $S$. The largest city in modern urban systems (the 'capital') is also special and often deviates from the overall systematic behaviour (Gabaix, 1999: Sect. V.1).

Under Gibrat's law, the probability density of the settlement sizes has a lognormal distribution rather than a power law (8). It has been argued that the sizes of large modern cities can be approximated by the lognormal distribution better than by a power-law (Eeckhout, 2004). The difference between the two distributions at large sizes is not very strong, so that it is difficult to draw this distinction confidently (Malevergne et al., 2011). It has also been pointed out that the rank–size analysis is often sensitive to the size of the smallest settlements, i.e., to the range of sizes considered.

Several explanations have been proposed for the origin of the power-law probability distribution in a settlement system that obeys Gibrat's law. Gabaix (1999) shows that a system with a fixed total number of settlements reaches an equilibrium state (and hence the probability distribution of suitably normalized settlement sizes no longer evolves) if the size of a settlement cannot be smaller than a certain (arbitrarily small) value. Then the equilibrium state has a power-law distribution with $\alpha = 1$, i.e., precisely Zipf's law. The size of each settlement varies in time systematically at a rate $\mu$ together with the total population. To make the probability distributions at different times comparable to each other, the equilibrium state is described in terms of the settlement sizes normalized to the total area $\tilde{S}(t)$ of all the settlements at a given time $t$. Within this model, deviations from Zipf's law arise because of deviations from Gibrat's law. Specifically, if the growth rate $\mu$ and its variance $\sigma^2$ depend on $S$, the exponent $\alpha$ in Equation (8) varies with $S$ as (Gabaix and Ioannides, 2004: 18)

$$\alpha(S) = 1 - 2\frac{\mu(S) - \bar{\mu}}{\sigma^2(S)} + \frac{S}{\sigma^2(S)}\frac{\partial \sigma^2(S)}{\partial S}, \tag{9}$$

where $\bar{\mu}$ is the growth rate averaged over all settlement sizes.

Reed (2001) relaxes the assumption that the number of settlements is fixed and his model allows for emergence of new settlement, e.g., by fission or otherwise. Then a power-law probability distribution follows from Gibrat's law because different settlements are founded at different times (see also Reed and Hughes, 2002b). This model is illustrated in Figure 2 where, for the sake of clarity, new settlements emerge at regular time intervals of 300 years rather than at random.

Suppose that the probability of a new settlement (perhaps via fission of villages into daughter communities – Duffy, 2015) to emerge in the time interval from $t$ to $t + dt$ is equal to $\lambda\,dt$, where $\lambda$ is the growth rate of the number of settlements. Then the expected (mean) number of settlements at a time $t$ is proportional to $\exp(\lambda t)$. All new settlements are assumed to be formed having the same size $S_0$. In fact, both birth and death of settlements is possible: then $\lambda$ is the difference between those rates (Gabaix, 2009: Sect. 3.4.2) The probability distribution of the settlement sizes at large times (later than about 800 years in the example of Figure 2) is a mixture of lognormal distributions. Reed (2001) shows that the resulting probability density has a power-law form at *both* small and large values of $S$ (a double Pareto probability distribution),

$$p(S) = \frac{1}{S_0} \frac{\alpha\beta}{\alpha+\beta} \left(\frac{S}{S_0}\right)^{-\alpha-1} \text{ for } S > S_0, \quad \text{and} \quad p(S) = \frac{1}{S_0} \frac{\alpha\beta}{\alpha+\beta} \left(\frac{S}{S_0}\right)^{\beta-1} \text{ for } S \leq S_0, \quad (10)$$

where $\alpha$ ($> 0$) and $\beta$ ($> 0$) are related to the model parameters $\mu, \sigma$ and $\lambda$:

$$\frac{\mu}{\sigma^2} = \frac{1}{2}(1 - \alpha + \beta) \quad \text{and} \quad \frac{\lambda}{\sigma^2} = \frac{1}{2}\alpha\beta. \quad (11)$$

In particular, when the growth rate of the number of settlements is equal to the population growth rate, $\lambda = \mu$, we have $\alpha = 1$ (Zipf's law) and $\beta = 2\mu/\sigma^2$. It is useful to derive the mean area of the settlements $\bar{S}$ in a system with the largest and smallest areas $S_{max}$ and $S_{min}$, respectively. For $\alpha \neq 1$, we obtain

$$\bar{S} = \int_{S_{min}}^{S_{max}} S p(S) \, dS = S_0 \frac{\alpha\beta}{\alpha+\beta} \left[ \frac{\alpha+\beta}{(\alpha-1)(\beta+1)} - \frac{(S_{max}/S_0)^{1-\alpha}}{\alpha-1} - \frac{(S_{min}/S_0)^{\beta+1}}{\beta+1} \right], \quad (12)$$

but a different result follows for $\alpha = 1$:

$$\bar{S} = S_0 \frac{\beta}{1+\beta} \left\{ \frac{1}{\beta+1} \left[ 1 - \left(\frac{S_{min}}{S_0}\right)^{\beta+1} \right] + \ln \frac{S_{max}}{S_0} \right\}.$$

Furthermore, $\bar{S}$ is undefined for $S_{max} \to \infty$ if $\alpha < 1$ since the integral in this equation diverges, but this fact should not prevent the calculation of the mean area within a finite range of $S$ using Equation (12) (or the one for $\alpha = 1$ as appropriate). Since $\alpha$ differs from unity sufficiently strongly in all the fits to the areas of the permanent Trypillian settlements discussed in Section 6, we used Equation (12) in the calculations reported below.

An additional feature added to the model by Reed (2002, 2003) is a random scatter in the initial settlement size $S_0$ (and a certain kind of its variation with time). If the probability distribution of the initial settlement sizes is lognormal, the resulting probability distribution of the settlement sizes at a later time is a mixture of lognormal distributions that the author calls the 'double Pareto-lognormal distribution' (dPlN). This probability distribution has power-law tails approximated by Equation (10) with a smooth transition between them (Reed and Jorgensen, 2004). Giesen et al. (2010) demonstrate that this distribution provides a better fit to modern city sizes than the lognormal distribution (see also Reed, 2001, 2002).

The explanation of Zipf's law as a consequence of Gibrat's law of proportionate growth and its modifications (Gabaix, 1999: Section V) relies on the description of the growth of settlements as a stochastic (random) process and does not appeal to any specific socio-economic processes. This may be considered a disadvantage and other approaches were developed assuming that the city sizes are determined by the requirement of maximum efficiency with respect to an interaction of various economic and social factors (e.g., Brakman et al., 1999; see Gabaix and Ioannides, 2004, for a review, and Duffy, 2015, for an archaeological context). However, the success of various stochastic models in explaining power-law behaviours in a broad variety of phenomena suggests that its nature is more general than any specific socio-economic mechanism. It is reasonable to consider social and economic factors as the determinants of the parameters of the stochastic models such as the mean growth rate $\mu$, the magnitude of its fluctuations $\sigma$, the rate of emergence of new settlements $\lambda$, etc. However, the overall features of settlement systems appear to be controlled by generic stochastic effects (Reed, 2001; Reed and Hughes, 2002b; Gabaix, 2009).

### *4.3. Settlement area and population*

A convenient and easily accessible measure of a modern settlement size is its population *N*. Reliable population data are usually available for modern systems but only rarely in prehistory. However, detailed analyses of modern population data demonstrate a close relation between the settlement area *S* and its population. Rozenfeld et al. (2011) use the UK and USA census data (1981 and 1991 for the UK and 2001 for the USA) to show that the population density *n* = *N/S* is nearly independent of the settlement area for settlements with population exceeding 5000 in the UK (83 percent of the total population) and 12000 in the USA (63 percent of the country's population). (These authors note a weak positive correlation between the population density and the city's population.) As a result, these authors find that Zipf's law holds equally well for the populations and the areas of the American and British settlements and cities. Sumner (1989) suggests that the population density in modern farming villages in the Fars Province of Iran has a typical value of 160 persons/ha, and argues that his value can be applied to archaeological settlements in a similar environment and under similar pre-industrial agricultural production and housing technologies. Duffy (2015) quotes ethnographic and archaeological evidence for diverse middle-range village systems where population density within the villages ranges from 30 to 500 persons/ha. The geometric mean for this range is about 120 persons/ha. Duffy (2014: 123–129) suggests a population density of 220 persons/ha at tell sites in the Bronze Age Great Hungarian Plain and 80 persons/ha outside the tell fortifications.

Population density independent of the settlement size implies that the socio-economic efficiency of human interactions is independent of the settlement size (constant returns to scale). If this is the case, the settlement population is simply related to its area, $N = nS$ with a constant population density *n*. However, Ortman et al. (2014, 2015) propose that the socioeconomic outputs increase with the scale of settlements more rapidly than their population (increasing returns to scale) and develop models for such an increase. These authors apply the model to modern and archaeological settlement data (the pre-Hispanic Basin of Mexico) and find a good agreement. In accordance with their model, the area of smaller settlements increases with its population not linearly but rather as $S = bN^{2/3}$, with $b \approx 0.2$–$0.3$ ha. The areas of larger settlements, which have structured and differentiated urban space with streets and other transportation networks, depend on the population size differently, as $S = b'N^{5/6}$ with $b' \approx 0.1$–$0.3$ ha. The borderline between the non-differentiated (amorphous) and urban (networked) settlements is at *N* = 5000 inhabitants in their data. Settlements less than 1 ha in area were excluded from the analysis due to poorer precision of the recorded archaeological data.

In the absence of any better estimates, we use the archaeological site area for the settlement area.

### *4.4. Settlement patterns in archaeology*

It is a truism to say that there is no certainty that even the largest archaeological settlements have all been discovered in any given case, and this can distort the rank–size dependence profoundly. In applications to modern urban systems, the sizes of the largest cities (of low ranks) are firmly known. Earlier archaeological rank–size analyses assign Rank 1 to the largest *known* settlement, discuss whether the observed rank–size dependence is convex or concave, and then interpret the result in terms of the centralization (or otherwise) of the settlement system (Johnson, 1977, 1980; Savage, 1997; Drennan and Peterson, 2004). Zipf's law as a special case of a power-law dependence of the form $R = kS^{-\alpha}$, with a certain exponent $\alpha$ ($\alpha = 1$ in Zipf's law). A power-law dependence is said to be *scale-free* meaning that it does not involve any special values of either *S* or *R*. In the present context, this implies that processes that cause the emergence of a power-law dependence of the settlement size on its rank must be independent of the settlement size (see Section 5). As a result, the regularity expressed in terms of a power-law, Zipf's law in particular, would not be affected if *all* the settlements of an area larger or smaller than any threshold had been removed from the data. However, the rank–size dependence can be distorted if *some* sites of large size remain undiscovered. Thus, the rank–size analysis relies heavily on the recovery rate of the less numerous settlements. The incompleteness of the data at intermediate sizes is even more damaging as this can cause spurious deviations from Zipf's law. The rank–size dependence refers to individual settlements, so, for instance, failing to include a few large (even if the largest one is included) settlements would produce a spurious primate (concave) rank–size distribution. If there are reasons to *assume* that Zipf's law applies at least within a limited range of settlement sizes, a conservative approach (as opposed to presuming that the data include all the higher-rank settlements) is to fit Zipf's law to that part of the rank–size dependence where it provides a reasonable approximation to the data and then to analyse any deviations from Zipf's law for the remaining settlements. This would not affect the fact that the observed distribution is convex or concave, but details of the interpretation may change. Savage (1997) addresses this problem assuming that the observed sample of settlements is drawn at random from a larger 'universe' of settlement that follow Zipf's law perfectly. This approach helps to decide whether the data are consistent with Zipf's law but does not help if any deviation has been detected.

The duration of the culture-typological stages, 300–700 years in the case of Trypillia, is a factor of ten larger than the expected lifetime of a settlement. Thus, unlike the case of modern cities where the data are known to be coeval, archaeological data provide a mixture of settlements that are unlikely to have all existed simultaneously (Dewar, 1991, 1994; Kintigh, 1994) forming an archaeological palimpsest (Duffy, 2015). Gabaix (1999: Section III.2) shows that combining *regional* (but simultaneous) data, where settlements in each region obey Zipf's law, leads to Zipf's law for the whole settlement system. His arguments apply to any other common power-law dependence. In other words, spatial variations in the settlement dynamics do not affect a power-law regularity if they do not affect locally the assumptions behind it (most importantly, Gibrat's law). It is not immediately obvious that Zipf's law is robust with respect to combining data from different time intervals when individual settlements emerge and disappear during the evolution (so that the total number of settlements can vary in time) but the temporal resolution of the data is insufficient to provide a picture of a system of *contemporaneous* settlements. Kohler and Varien (2010) carefully use data from narrow time intervals of 20–50 years in their rank–size analysis. However, dating precision at this level is an exception rather than a rule. When settlements can both form at a rate $\lambda_1$ and die at a rate $\lambda_2$, and the two processes have similar statistical properties, the above arguments appear to be applicable with $\lambda = \lambda_1 - \lambda_2$. Reed and Hughes (2002a,b) discuss such birth and death processes. Reed (2002: Section 4) shows that the dPlN model applies also when the new settlements are formed only for a limited period of time rather than during the whole evolution of the settlement system. This model is robust with respect to variations in the model of settlement foundation process, for example, when the rate of growth varies and the evolution even includes periods of decay: what is important is that Gibrat's law of proportional effects remains applicable (Reed, 2002: 13). The system of Trypillian settlements is of this type: the rate of its growth is variable and the system decays towards the end of the evolution.

It can be argued that the relatively long duration of the culture-typological phases may not affect the approach suggested here too strongly. Suppose that the probability density of the settlement sizes, for instance that given in Equation (10), varies with time, e.g., because the parameters $S_0, \alpha$ and $\beta$ are variable. Consider the case when the temporal resolution of the settlement data (the duration of a culture-typological phase in our case, starting at time $t_1$ and ending at $t_2$) is larger than the lifetime of a settlement, as happens in our case. Then the quantity obtained from the archaeological data is the cumulative probability over the duration of the resolved time interval $t_1 \leq t \leq t_2$, and then an appropriate quantity can be the probability density per unit time interval,

$$\tilde{p}(S) = \frac{1}{\Delta t} \int_{t_1}^{t_2} p(S, t)\, \mathrm{d}t,$$

where $\Delta t = t_2 - t_1$ is the duration of the time interval. The dependence of $\tilde{p}(S)$ then remains the same as that of $p(S, t)$ if the latter represents the product of two functions, one depending on $S$ alone and the other of $t$ alone. This condition may be too restrictive, but then the values of $S_0, \alpha$ and $\beta$ obtained below should be understood as effective values that characterise the phase as a whole.

The settlement area is not necessarily equal to the archaeological site area. Geophysical prospecting, aerial surveys, surface inspection and archaeological test pits are bound to produce different estimates when applied to similar settlements. Even with geophysical prospecting, the measured area is likely to be a lower estimate of the actual settlement area because not all parts of the settlement might have left a detectable signal.

Modern cities follow Zipf's law with a striking universality and precision (Gabaix and Ioannides, 2004 and references therein). This can be explained by the fact that city sizes are expected to converge to a power-law equilibrium at a time scale of order hundred years (Gabaix, 1999: 741) or that the birth of new settlements produces a power law distribution over a similar time scale (Reed, 2001, 2002). The resulting probability distributions of the settlement size are rather independent of the specific nature of the processes that control the settlement system if only they satisfy general requirements of probabilistic nature. Many socio-economic, physical, astronomical, biological, linguistic, etc., phenomena exhibit similar power-law regularities (e.g., Reed and Hughes, 2002b; Gabaix, 2009), consistent with the fact that a power-law distribution is an outcome of many random growth processes. Duffy (2015) suggests six processes that can affect site size hierarchies in stateless farming societies. When a sufficiently large number of such independent processes are in action together, the outcome is likely to be as good as random.

In the model of Gabaix (1999: Section V.1), some deviations from Gibrat's law remain consistent with Zipf's law, but some cause deviations from it, e.g., the dependence of the settlement growth rate on its size (see also Gabaix and Ioannides, 2004: Section 3.2). For example, the largest settlement may be a peculiar phenomenon, with an outlier (the 'capital') being larger than what Zipf's law would predict for its rank; then fitting Zipf's law to smaller settlements seems to be a better option. Secondly, smaller settlements can deviate from Gibrat's law as their growth can fluctuate stronger (both in time and among such settlements) than that of larger units because they lack the economic and social diversity and, hence, the robustness of the larger sites. As a result, the variation of the settlement size on its rank becomes flatter than Zipf's dependence as shown in Equation (9).

It is surprising that archaeological data usually deviate from Zipf's law rather strongly, provoking discussions of the social and cultural implications of the deviations: "Almost no data set corresponds exactly to the rank–size rule, so interpretations are based on how the data set diverges from the expected results" (Savage, 1997). Such deviations are often claimed to persist over centuries or even millennia. However, any discussion of such deviations should first identify reasons as to why the system explored would remain far from the equilibrium or deviate systematically from Gibrat's law at the time scale involved. This is done rarely if ever. We feel that the incompleteness of both the archaeological data and their interpretation may often be a more plausible explanation of the claimed deviations from Zipf's law.

With these caveats in mind, we proceed to the analysis of the Trypillian settlements.

# 5. Regularities in the Trypillian settlement system

Since the populations of Trypillian settlements are not known, we apply relations of Section 5 to the areas of archaeological sites selecting permanent settlements alone. Basic parameters of their areas are given in Table 1.

## 5.1. The number and total area of the settlements

As a first step, it is desirable to check if the data are consistent with the assumptions behind Gibrat's law. Given the paucity of the data, particularly their incomplete recovery (less than a half of permanent settlements have an area estimate of any accuracy, see Table 1), we cannot do much in this respect. However, we can test the assumption that the evolution of the system of settlements exhibits a systematic exponential growth. Figure 3a shows the total area of the permanent settlements versus time, as given in the last line of Table 1. With a linear scale on the time axis and logarithmic scale on the area axis, an exponential growth of the total area $\tilde{S}$, of the form

$$\tilde{S} = \tilde{S}_0 \exp(\mu \Delta t), \qquad (13)$$

would produce a straight line of the slope $\mu$, where $\tilde{S}_0$ is the initial total area, $\mu$ is the growth rate and $\Delta t$ is the time elapsed since the conventional start of the evolution. Overall, the data fit this dependence fairly well, at least better than might be expected from the fact that less than half of the settlements have a known area. However, the total area of Trypillia CII settlements does not fit this dependence being significantly smaller than that of the few preceding stages. This is also true of the BII-CI but to a lesser extent, so that the deviation of this stage from the general trend is comparable to the general scatter of the other data points. Therefore, we fitted an exponential function of the form (13) to the Trypillia A, BI, BI-II, BII, BII-CI and CI areas. More precisely, we fitted with linear least squares the dependence of ln $S$ on the time lag $\Delta t$ between the median date of each stage and 5100 BC, the median date of Trypillia A. The result, shown with solid line in Figure 3a, has $\tilde{S}_0 \approx 67 \pm 9$ ha and the area growth rate $\mu = 1/(460 \pm 100 \text{ years})$, where the ranges indicate the standard error. In other words, the total area of the settlements was increasing by a factor of two every $\mu^{-1} \ln 2 = 320 \pm 70$ years, starting at 60–70 ha in Trypillia A. This value of the initial area is comfortably close to the measured total initial area of 61 ha for Trypillia A in Table 1. We stress again that these are the areas of only those settlements that have known area; the total area of permanent settlements in the system is likely to be twice as large if the areas of the settlements with unknown areas are similar to those available.

    Figure 3b shows, in a similar manner, the evolution of the *number* of permanent settlements where we show the data for both all permanent settlements (circles, red) and for those with known areas (triangles, blue). These data are taken from the fourth and fifth lines of Table 1. As with the areas, the number of settlements increases systematically (albeit with significant scatter around the fit line) and Trypillia CII is again different from the other stages having a disproportionately small number of settlements, either all or those with known areas. As with the areas, the fitted exponential dependences of the number of settlements $M$ on the time lag, $M = M_0 \exp(\lambda \Delta t)$, are shown solid and dashed. The total number of permanent settlements grows with time at a rate $\lambda = 1/(1020 \pm 320 \text{ years})$ (which corresponds to the doubling time $\lambda^{-1} \ln 2 = 710 \pm 220$ years), with $M_0 = 38 \pm 4$ the number at the middle of Trypillia A. Similar estimates for the sites of known area are somewhat different at $\lambda' = 1/(540 \pm 330 \text{ years})$ and $M'_0 = 14 \pm 3$ (the doubling time $\ln(2)/\lambda' = 370 \pm 230$ years). The difference between $\lambda$ and $\lambda'$ reflects an increasing fraction of the settlements with known area among the permanent settlements discovered; the growth rate of the total number of settlements, $\lambda$, is a more useful quantity. It is important to remember the existence of this difference when interpreting results obtained below from the settlement areas: they may be affected by the incompleteness of the data.

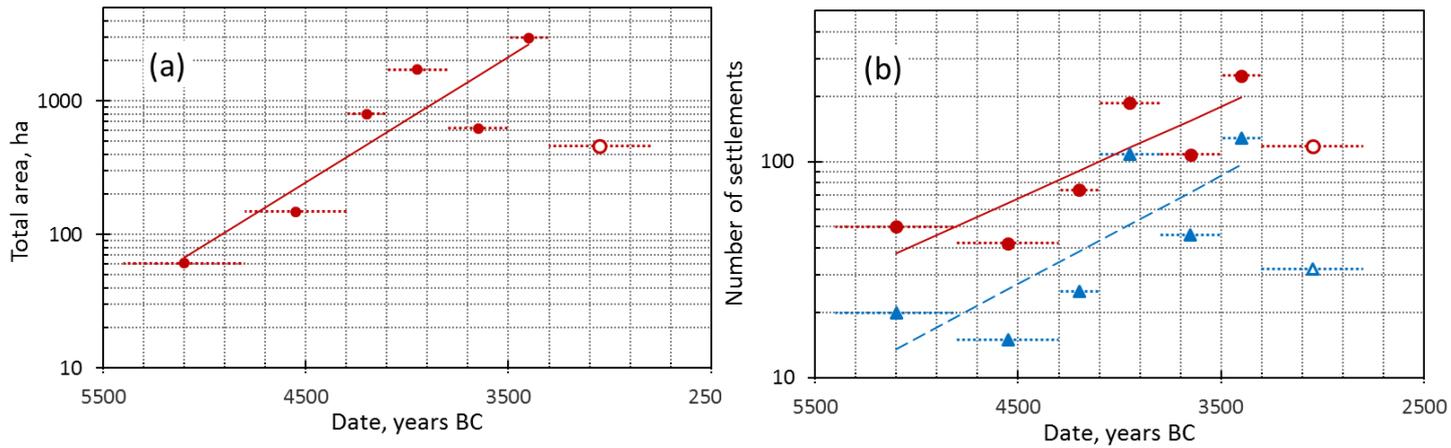

Figure 3. The evolution of (a) the total area and (b) the number of permanent Trypillian settlements. Each data point (circle or triangle) is assigned to the median date of the corresponding typo-chronological stage, and horizontal dotted lines show the duration of each stage as given in Table 1. Solid line in Panel (a) shows the exponential fit to the total areas of Trypillia A, BI, BI-BII, BII, BII-CI and CI. The area of Trypillia CII settlements, shown with open circle, has been excluded from the fitting. In Panel (b), circles show the total number of permanent settlements whereas triangles are for those with known areas. The exponential fits to them are shown with the solid (red) and dashed (blue) line, respectively. Stage CII is excluded from the fitting and shown with open symbols.

Thus, the number of settlements appears to grow somewhat faster than their total area (although the fitted values of the growth rates $\mu$ and $\lambda$ are in a marginal agreement within two standard deviations) implying that new settlements were usually small and grew in size significantly before being abandoned.

The deviation of Trypillia CII from a systematic, approximately exponential growth of the total area and number of the settlements is understandable as this is the terminal stage of the Cucuteni–Trypillia complex characterized by the decay of the settlement system.

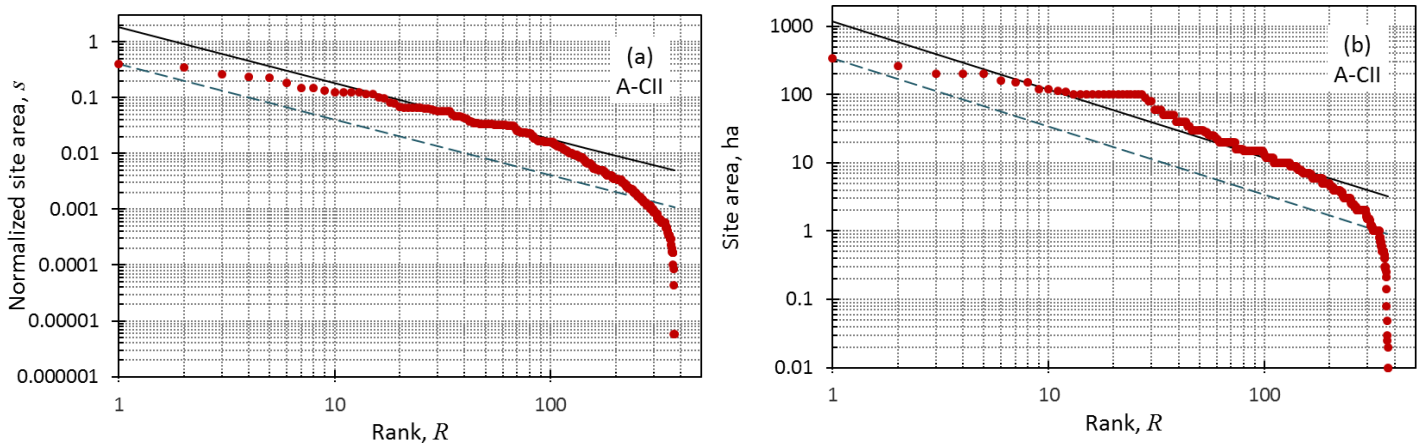

Figure 4. The rank–size distribution of the normalized (**a**) and non-normalized (**b**) areas of permanent settlements from Trypillia A–CII (circles) and Zipf's law $s = aR^{-1}$ with different vales of the normalization factor $a$ (solid and dashed). With the logarithmic scales on both axes, a power law of the form $s = aR^{-\alpha}$ is represented by a straight line of the slope $-\alpha$. In Panel (a), the settlement areas in each stage are normalized to the total known settlement area in that stage shown in the bottom line of Table 1. Zipf's law, $\alpha = 1$, shown with solid line is normalized to fit the part of the data that is closest to a power law, $a \approx 1.845$. The dashed line only differs from the solid line in its normalization to fit the area of the settlement of Rank 1, $a \approx 0.405$.

## 5.2. Cumulative settlement patterns

Before considering individual Trypillia stages, it is useful to explore the data for all the stages combined: with a larger number of settlements, it may be easier to detect any systematic and interesting features which can be sought for in the data for individual stages. The rank–area distribution of the known areas of permanent settlements from all Trypillia stages is shown in Figure 4a. To compensate for the systematic increase in the total area of the settlements, we have normalized the area of each settlement by the total area of the permanent settlements of its stage, so that the normalized area *s* is given by $s = S/\tilde{S}$, where $S$ is the site area and $\tilde{S}$ is the total area given for each stage in the last line of Table 1. Excluding the Trypillia CII sites does not affect the plot much; it only makes the tail at small values of *s* shorter: a relatively small number of permanent settlements of this stage have the area known, and most of them are small. The straight lines in Figure 4a represent Zipf's law fitted either to those settlements ($13 < R < 107$, $0.125 > s > 0.014$) where it appears to be a fair approximation to the data (solid line) or to the single point of Rank 1 which is the usual practice in the archaeological literature (dashed line).

The effect of the normalization of the site areas can be appreciated by comparing Figure 4a with Figure 4b showing the rank–size plot for the absolute rather than normalized areas. The difference in the shape of the dependence is noticeable although not dramatic. In particular, the two plots appear to deviate from Zipf's law to different extents. A deviation from the power-law trend at $13 < R < 27$ in Figure 4b, where *S* is exactly equal to 100 ha for a significant number of sites, is clearly due to the tentative nature of many area measurements.

Strong deviations from Zipf's law at high ranks occur in many modern and archaeological data sets (e.g., Johnson, 1980; Savage, 1997; Drennan and Peterson, 2004; Griffin, 2011); we discuss these deviations later. The rank–size distributions of Figure 4 deviate from Zipf's law at smaller ranks, exactly where it should be most accurate. However, the character of the deviation depends on whether Zipf's law is normalized to fit the data at medium ranks or at $R = 1$. In the former case, shown in Figure 4 with solid line, one would conclude that the deviation occurs at smaller ranks $R < 10$ (larger sites), with the biggest settlements being smaller than expected from Zipf's law. If, however, we were confident that the largest Trypillia settlement (at present, Talianki of Trypillia CI, 341 ha in area) has been found, the correct normalization of Zipf's law would be that shown in Figure 4 with dashed line. Then we would have concluded that the convex shape of the rank–size distribution suggests a decentralized system of settlements with medium-size settlements larger than expected. However, we believe that neither of these conclusions is viable for the following reasons.

## 5.2.1. The probability distribution of the settlement areas

A disadvantage of the rank–size representation of the settlement system is that is attaches disproportionate importance to a few largest sites. This is acceptable in the case of modern data, where the sizes of the largest cities are known confidently, but becomes problematic in the case of archaeological data. For instance, if just a few larger settlements were not found, a perfect Zipf's distribution would be transformed into a primate one where the settlement areas apparently decrease with their rank faster than in Zipf's law. Among Trypillian settlements, the ten biggest settlements have the areas ranging from 341 to 120 ha. Their estimated areas (mostly multiples of 100 or 10 ha) rightly suggest that the area estimates are rather crude and, hence, the rank–size distribution is uncertain, especially at low ranks where the number of settlements is small. Moreover, when settlements of some sizes (especially very large or very small) do not occur in the data or are found only once, binning is required to obtain reliable results. Malevergne et al. (2013: Appendix A.3) show how a limited size of the sample of settlements can damage the rank–size dependence at its large-size end.

Working with the probability density $p(s)$ helps to avoid this difficulty, at least in part. Furthermore, the use of the probability density instead of the rank–size distribution is better consistent with theoretical approaches to the settlement patterns (Section 5). The simplest way to estimate the probability density is by binning the data into discrete area ranges. The number of settlements in a bin (the 'frequency') with the area $S$ is approximately equal to $Mp(S)$, where $M$ is the total number of settlements. Binning leads to a loss of information, so that narrower bins are preferable. However, their number is limited by the amount of data available. To obtain stable results and to avoid misleading graphical representations, the value of $Mp(S)$ should be large enough in all the bins (as a rule of thumb, the expected frequency should be at least five in each bin in 80 percent of the bins – Siegel and Castellan, 1988: 49).

Figure 5 shows the number of settlements per bin width of 0.02 in terms of the normalised areas, together with the fitted Zipf's law. When presented in this form, the data do not exhibit any apparent deviations from Zipf's law apart from a random scatter.

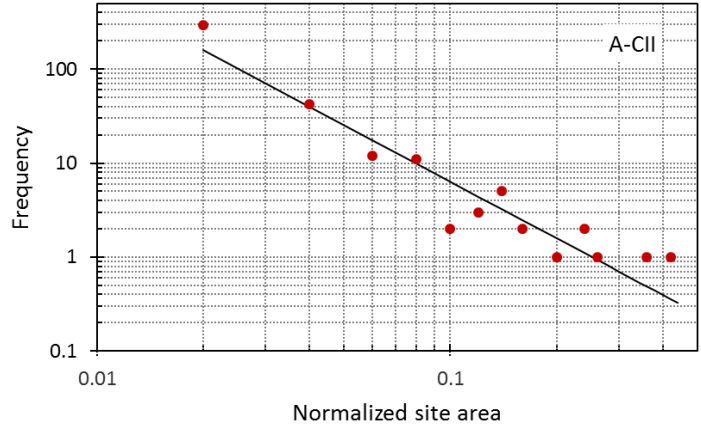

Figure 5. The number of settlements per bin of 0.02 in width versus the normalized settlement area. As in all similar figures shown below, the area $s$ used to identify a bin is the largest value within the bin. Data for all the Trypillia Stages A–CII are shown with red circles ($M = 374$), and Zipf's law is shown with solid line.

In fact, binning all the small settlements into a single bin, with 291 sites with $s \leq 0.2$, hides their interesting and informative behaviour. A remedy, well known in physics, is to use area intervals (bins) whose width increases with $s$. This presentation also helps to avoid low frequencies at larger values of $s$ as the decrease in the probability with area is then compensated by the increase in the bin width. For a power-law form of $p(s)$, the expected number of data points per bin remains constant for all values of $s$ when the bin width $\Delta s_i$ increases logarithmically as $\ln(s)$ or, equivalently, as $\Delta s_i \propto n^i$, where $i = 0, 1, 2, \ldots$ is the bin number and $n$ is any integer number chosen as convenient. With the variable bin width, the form of the probability density obtained from the binned data is slightly modified. If $p(s)\,ds$ is the probability of a measurement to be in a bin of a width $ds$ centred at $s$, with logarithmic bin widths we have a modified form $\tilde{p}(s)$, such that $p(s)\,ds = \tilde{p}(s)\,d\ln(s)$, so the probability density corresponding to Zipf's law is now given by, instead of Equation (8),

$$\tilde{p}(s) = as^{-\alpha}.$$

The normalized area distribution of permanent Trypillia A–CII settlements with bins of variable width $\Delta s_i = 2^i \times 10^{-5}$ is shown in Figure 6. Solid line in Figure 6 shows this dependence with $\alpha \approx 1$, Zipf's law for large settlements. It is remarkable that the small settlements have now been resolved into a dependence different from Zipf's law and consistent with the predictions of Reed (2001, 2002) and Reed and Jorgensen (2004). The power-law fit as small values of $s$ shown with dashed line in Figure 6 has the form

$$\tilde{p}(s) \propto s^\beta, \quad \beta \approx 1.78.$$

We restrain ourselves from fitting the continuous dPlN probability distribution of Reed (2002), valid at any value of $s$, and applying any advanced statistical fitting procedures and tests for the quality of the fit. The data available are too scarce and approximate to warrant such a detailed analysis. Furthermore, although the cumulative data for all the stages, shown in Figure 6 and Figure 7, suggest the dPlN distribution (with the data points that are shown with open circles providing a smooth transition between the two asymptotic power laws), the probability distributions for individual stages discussed in Section 6 appear to be consistent with the simpler double Pareto distribution. This may imply that the random scatter in the size of the newly formed settlements was insignificant within the time span of each stage and only becomes important over longer time spans.

Thus, we fit the power-law tails to the probability distributions of the site areas,

$$\ln p(S) = y_\alpha - \alpha \ln S \text{ at large } S \quad \text{and} \quad \ln p(S) = y_\beta + \beta \ln S \text{ at small } S, \qquad (14)$$

where $y_\alpha, \alpha, y_\beta$ and $\beta$ are the fitted parameters and $S$ is measured in hectares (for the normalized areas, $S$ is replaced by the dimensionless area $s$). These fits can be interpreted in terms of the probability distribution (10) and (11). The results are illustrated in figures below and summarized in Table 2. It is not quite satisfactory that some fits are obtained for just a handful of data points. However, the bin widths used are already rather narrow and making them even narrower would lead to an unacceptably large number of bins with a small number of measurements. However, the variation of the fitted parameters between various Trypillia stages does not appear to be inexplicable or random (see below) and thus deserves attention. In addition, the quality of the fits is rather high: the values of $R^2$ (see the caption of Table 2) are close to unity, and the $\chi^2$ test applied following Siegel and Castellan (1988: 45) is safely satisfied for all the fits reported. We also note that our aim is to fit a specific model to the data rather than to select the best model among many. This provides additional justification for our fits based on a small number of data points.

We first consider the cumulative normalized areas of Trypillia A–CII shown in Figure 6. Fits of the form (14) in the ranges $0.02048 < s \leq 0.65536$ and $10^{-5} < s \leq 6.4 \times 10^{-4}$ (five and four bins in each range, respectively) are shown solid and dashed; they have the slopes $\alpha \approx 1.12$ at larger values of $s$ and $\beta \approx 1.78$ at smaller areas; the two fits intersect at $s_0 \approx 0.0035$.

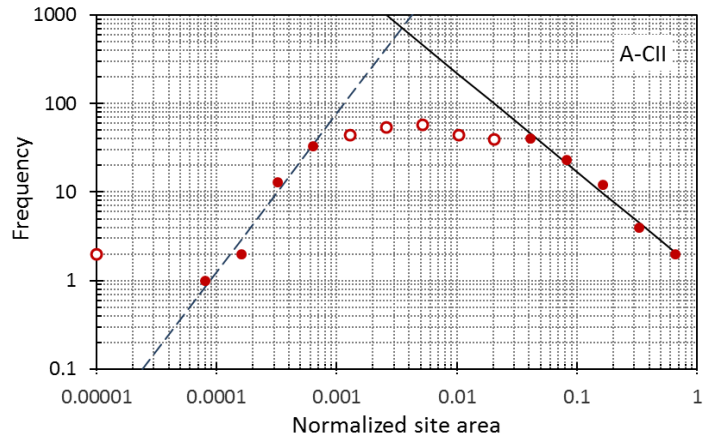

Figure 6. As in Figure 5 but with logarithmic bin widths, $\Delta s_i = 2^i \times 10^{-5}$ ($i = 0, 1, 2, 3, \ldots$). The straight lines represent the least-squares fits to the data points shown with filled circles, $s^{-1.1}$ at large $s$ and $s^{1.8}$ at small $s$. The data points shown with open circles have been excluded from the fits.

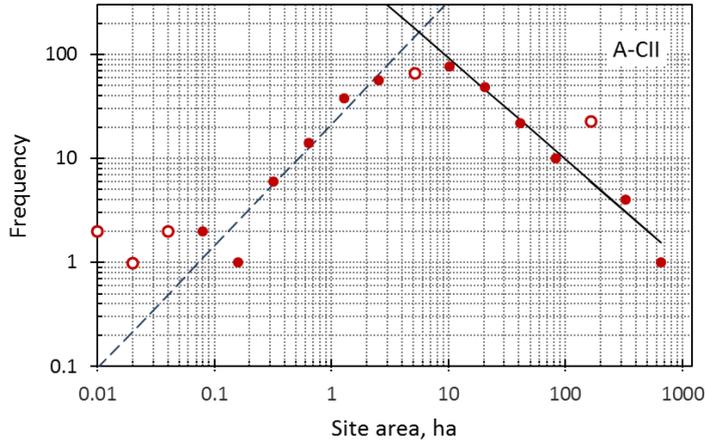

Figure 7. As in Figure 6, but for absolute settlement areas with bin widths of $\Delta S_i$ [ha] $= 2^i/100$ ($i = 0, 1, \ldots$).

For comparison, we show in Figure 7 the estimated probability density of the absolute settlement areas per logarithmic area bin. The general features of the distribution are similar to those of the normalized areas although the scatter is stronger. Solid line fits the data in the bins at $5.12 < S \leq 655.36$ ha excluding one bin at $81.92 < S \leq 163.84$ ha; the fitted slope is $\alpha \approx 0.98$. The dashed line is fitted at $0.04 < S \leq 2.56$ ha, its slope is $\beta \approx 1.18$, significantly smaller than that in the fit to the normalized areas. The two fits intersect at $S \approx 5.7$ ha.

## 6. Evolution of Trypillian settlements

The evolution of the system of settlements can only be considered between the archaeological stages as more precise dating is not available. Within each stage, the fitting results are the same for either normalized or absolute settlement areas. Therefore, we use the absolute areas measured in hectares. The frequency distribution of the permanent settlement areas are shown in Figure 8 for the individual stages together with the power-law fits. In all cases, a power-law is a reasonable form to fit the data at both small and large values of $S$. The data points involved in the fits are shown with filled circles, whereas those that do not conform to them are represented by open circles: these data were not used in the fitting. The fits are shown with solid lines at large $S$ and dashed at small $S$. Detailed parameters of the fits are presented in Table 2 using the notation of Equations (14) and (10). Equation (11) was used to estimate $\mu/\sigma^2$ and $\lambda/\sigma^2$. We attempted at improving the statistical quality of the fits by combining data from several stages. The results are shown in Figure 9 and Table 3. We present rounded values of the parameters and their standard deviations in the tables, but all calculations have been done with exact values.

The tables also contain quantities that characterise the quality of the fits. The parameter denoted $F$ with the corresponding subscript ($F_\alpha$ or $F_\beta$ at large and small areas, respectively) characterizes the significance of the model parameters ($\alpha$ and $y_\alpha$ or $\beta$ and $y_\beta$, respectively): if $F$ exceeds 0.05, the admissible range of at least one of the fitted parameters includes zero with the 95% confidence, so that that parameter is said to be insignificant in that sense that its inclusion into the fitted model is not justified by the data. A measure of the quality of the fit is provided by the value of $R^2$ (the coefficient of determination) that quantifies the fraction of the variation in the data accounted for by the fit; $R^2 = 1$ corresponds to a perfect fit. The values of $R^2$ are rather high for all the fits reported below ($R^2 > 0.7$) except for Stage CII. We also note that the fit residuals do not show any evidence of any systematic trends additional to the power-law variations fitted.

Figure 8 shows little evidence of a smooth transition between the power-law asymptotics at small and large values of $S$, except perhaps in Stages BII and CI where the data are more abundant. For this reason, we fitted the simpler asymptotic forms of Equation (10) or (14) rather than the more general dPlN distribution.

There are only 20 permanent settlements with known areas in Trypillia A and many bins are empty. In particular, there are only three non-empty bins at $S \leq 2.56$ ha, so that the fit at small $S$ relies on a single site of 0.03 ha in area (Chernyatka-Shumyliv in the Vinnitsa Province). The fit at large $S$ appears to be more reliable. The data for Stage BI have similar deficiencies. Correspondingly, the fits have low significance, $F_\beta = 0.14$ and 0.06 for Stages A and BI, respectively. For Stage BI, the fit at large $S$ is also unsatisfactory with $F_\alpha = 0.1$. Thus, we also considered the combined data set of Stages A and BI: we sacrifice the temporal resolution of the results for their statistical quality.

The combined Stage A and BI data and the fits are shown in Figure 9 and Table 3. Even with the combined data, the number of non-empty bins and the number of measurements in a half of them are small, and the fit at small $S$ remains unsatisfactory with $F_\beta = 0.14$. Therefore, these fits should also be considered with caution.

Data paucity remains a problem for Trypillia BI-BII as well. In particular, the fit at large $S$ has low significance, $F_\alpha = 0.08$. A notable feature of this stage are the six largest settlements (out of 25 permanent settlements with known area) that clearly stand out and do not conform to the power-law pattern: Trypillia-Lypove (120 ha) in Kyiv and Veselyi Kut (150 ha), Bagachivka II, Khar'kivka, Myropillya and Vil'khovets II (100 ha each), all in Cherkassy Province. The next largest site has a significantly smaller area of 30 ha. We also considered combined data for Stages A, BI and BI-BII shown in Figure 9 and Table 3. The statistical quality of the fit to the combined data set is satisfactory but the time span of the combined stages is as large as 1300 years, and it is likely that the settlement system was changing significantly during this period. A worrying feature of the fits to the combined data sets is that the values of $\beta$ and $y_\beta$ differ significantly from those obtained for the individual phases. Combining the data does not lead to more reliable results.

There are only three non-empty bins at $S < 2$ ha in Stage BII and, as a result, the significance of the fit is low at $F_\beta = 0.1$. However, there are abundant data at large areas. On the contrary, the BII-CI data suffer from a small number of settlements at large areas but provide confident results at small values of $S$. Since the probability distributions of Stages BII-CI and CI are somewhat similar, we also considered their combined data shown in Figure 9. Results from the combined data appear to be satisfactory as they are consistent with those from the individual phases.

The data for Stages CI and BII are the best among the stages, with a large number of settlements. In particular, all areas are well represented in Stage CI. Both BII and CI data show signs of a smooth transition between large and small areas consistent with the dPlN model of Reed (2002, 2003) where the newly formed settlements have a range of areas rather than the unique area $S_0$.

Trypillia CII has clearly different properties from the other stages. The flat distribution of the site areas (relatively small values of $\alpha$ and $\beta$) suggest lower growth rates of both the area and the number of settlements. This provides a formal justification for excluding this stage from the exponential fits of Section 5.1. The data from this final stage of the Trypillia have a large scatter, especially at small areas; as a result, the coefficient of determination is as low as $R^2 = 0.56$.

A notable feature of the data is that all the data points at large areas fit well a power law, except for one point at BI-BII. This suggests that the mega-sites are ordinary members of the settlement system rather than an exceptional phenomenon at all stages except for BI-BII. In particular, the giant settlements of Stage CI, Maidanetske (about 200 ha) and Talianki (341 ha) fit the overall pattern. This does not mean that the role of the latter two settlements in the socio-economic system was insignificant but strongly suggests that they are a result of a normal settlement growth and conglomeration rather than of any exceptional processes. On the other hand, the large settlements of Trypillia BI-BII, if attributed correctly to this phase, appear to be truly exceptional and their origin deserves further analysis.

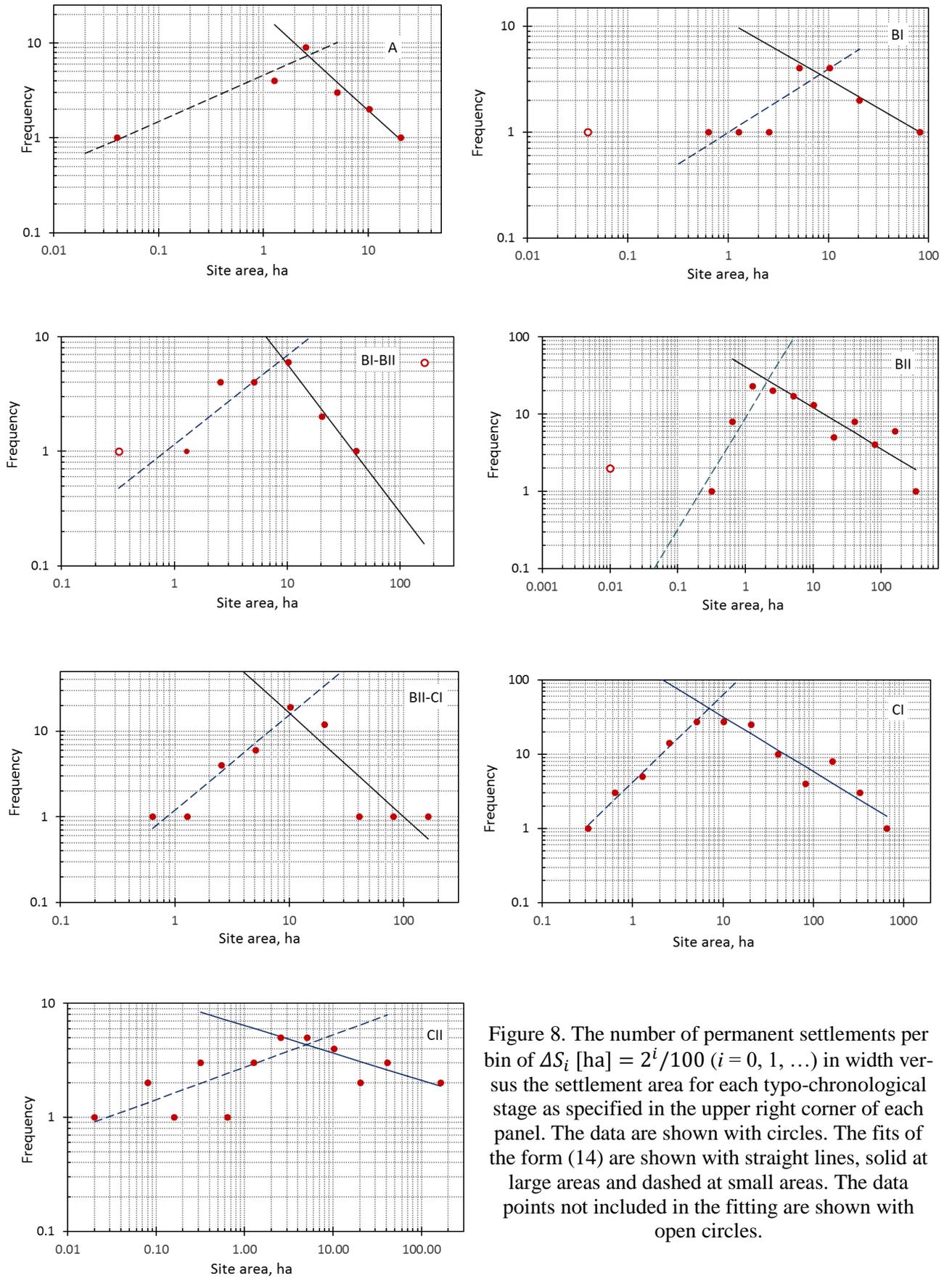

Figure 8. The number of permanent settlements per bin of $\Delta S_i$ [ha] $= 2^i/100$ ($i = 0, 1, \ldots$) in width versus the settlement area for each typo-chronological stage as specified in the upper right corner of each panel. The data are shown with circles. The fits of the form (14) are shown with straight lines, solid at large areas and dashed at small areas. The data points not included in the fitting are shown with open circles.

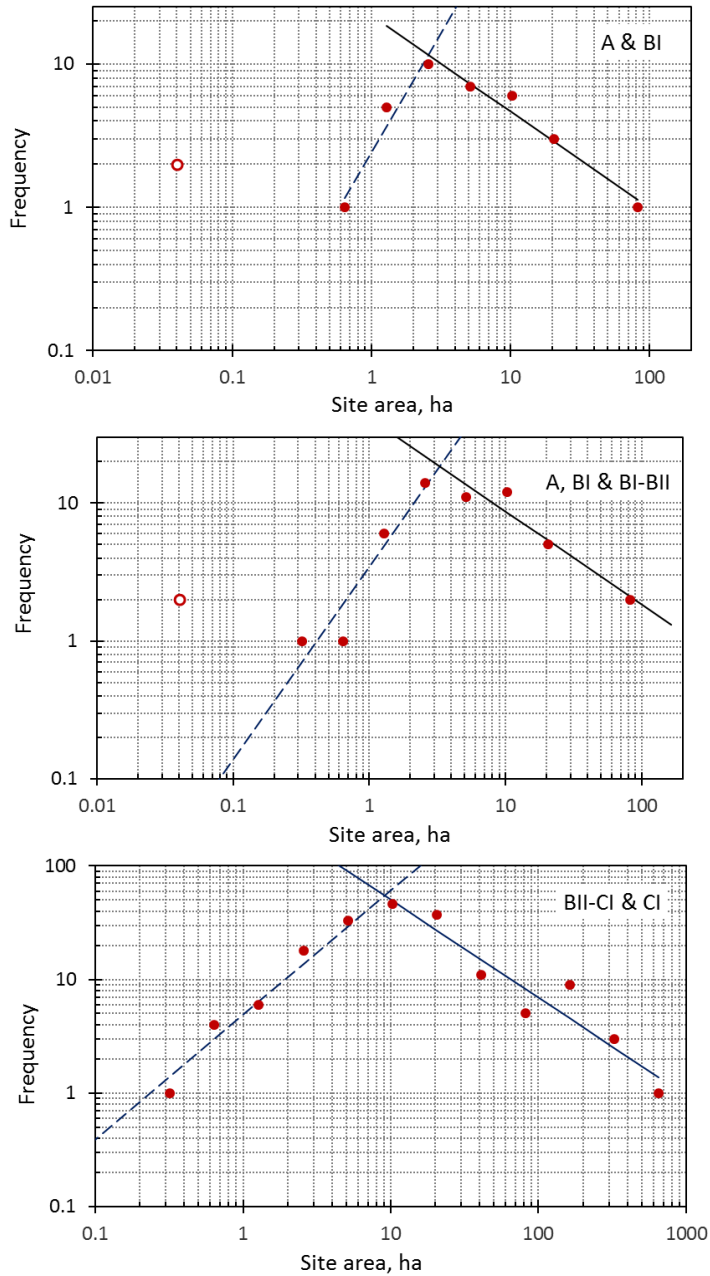

Figure 9. As Figure 8 but for combined stages as indicated in the upper right corner of each panel.

The statistical quality of some fits (whole Stages BI and BI-BII and at small areas in Stages A and BII) is not satisfactory as the values of $F_\alpha$ and/or $F_\beta$ are too large. However, our attempts to improve the statistics by combining the stages, reflected in Table 3, are not quite successful. For example, the mean values of $S_0$ obtained for Stages A and BI separately differ by a factor of 2–3, whereas $S_0$ from the combined data set is very similar to that for Trypillia A but not BI. The values of $\mu/\sigma^2$ and $\lambda/\sigma^2$ obtained from the combined data set are entirely different from those obtained for the individual Stages A and BI. The result of combining A, BI and BI-BII is similarly deficient. The values of these parameters inferred for Stages BII-CI and CI individually are close to each other, so their combined data yield the values of $S_0$, $\mu/\sigma^2$ and $\lambda/\sigma^2$ that are close to the average of the single-stage values. When the stages that are combined have different probability distributions of the settlement areas, a fit to their combination remains of poor quality. The loss of the temporal resolution is not a sufficient justification for combining the stages even in this case. We conclude that the settlement system evolved significantly from Stage A to Stage BII, and the resulting heterogeneous data cannot be meaningfully combined and analysed together.

# 7. Interpretation of the temporal patterns

Our interpretation of the evolution of the Trypillia settlements is based on Equations (10) and (11). The estimates of the growth rates of the settlement area, $\mu$, and their number, $\lambda$, both relative to the variance (volatility) of the area growth $\sigma^2$, are given for individual stages in Table 2 and plotted in figures below. This table also presents the estimated initial area of the newly founded settlements in each stage, $S_0$ obtained from Equations (10) as the area where the two asymptotic forms of $p(S)$ intersect:

$$S_0 = \exp\left(\frac{y_\alpha - y_\beta}{\alpha + \beta}\right). \quad (15)$$

In this section, we compare these model results to the observations to verify the model and to interpret the results.

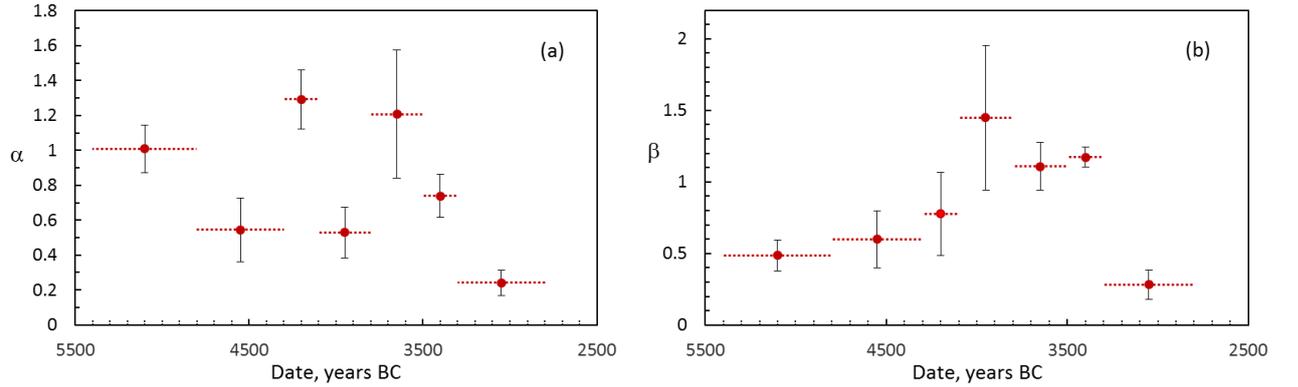

Figure 10. The fitted slopes $\alpha$ and $\beta$ of the asymptotic dependencies given in Equation (10) and shown in Table 2. The vertical error bars represent the standard deviation whereas the horizontal dotted lines indicate the duration of each stage.

Table 2. Parameters of the Trypillian settlement system obtained from least-squares fits of the form (14). The quantities shown, together with their standard deviations, are the power-law exponent $\alpha$ at large areas ($\alpha = 1$ for Zipf's law) and the intercept $y_\alpha$ at large areas, together with the corresponding significance value $F_\alpha$ and a measure of the overall quality of the fit $R^2$. The similar quantities at small areas, $\beta, y_\beta, F_\beta$, and $R^2$, are shown in the next three rows. The next four rows present the quantities derived from the fits: the area of the newly formed settlements $S_0$ from Equation (15), the values of $\mu/\sigma^2$ and $\lambda/\sigma^2$ of the stochastic growth model of Reed (2001) from Equation (11), and the mean area from Equation (12). The median and mean areas from the original data in Table 2 are given in the last two rows for comparison with $S_0$ and $\bar{S}$, respectively. For the reader's convenience, the values of $F$ exceeding 0.05 are shown underlined: these fits are not statistically satisfactory.

| Stage | A | BI | BI-BII | BII | BII-CI | CI | CII |
|---|---|---|---|---|---|---|---|
| $\alpha$ | $1.0 \pm 0.1$ | $0.5 \pm 0.2$ | $1.3 \pm 0.2$ | $0.5 \pm 0.1$ | $1.2 \pm 0.4$ | $0.7 \pm 0.1$ | $0.2 \pm 0.1$ |
| $y_\alpha$ | $3.0 \pm 0.3$ | $2.4 \pm 0.5$ | $4.7 \pm 0.5$ | $3.7 \pm 0.6$ | $5.6 \pm 1.4$ | $5.2 \pm 0.6$ | $1.9 \pm 0.2$ |
| $F_\alpha/R^2$ | 0.02/0.96 | <u>0.1</u>/0.82 | <u>0.08</u>/0.98 | 0.01/0.73 | 0.05/0.78 | 0.002/0.88 | 0.03/0.73 |
| $\beta$ | $0.5 \pm 0.1$ | $0.6 \pm 0.2$ | $0.8 \pm 0.3$ | $1.4 \pm 0.5$ | $1.1 \pm 0.2$ | $1.2 \pm 0.1$ | $0.3 \pm 0.1$ |
| $y_\beta$ | $1.5 \pm 0.2$ | $0.0 \pm 0.3$ | $0.5 \pm 0.2$ | $2.2 \pm 0.4$ | $0.2 \pm 0.2$ | $1.4 \pm 0.1$ | $1.0 \pm 0.2$ |
| $F_\beta/R^2$ | <u>0.14</u>/0.95 | <u>0.06</u>/0.75 | <u>0.12</u>/0.78 | <u>0.10</u>/0.80 | 0.007/0.94 | 0.0004/0.99 | 0.03/0.56 |
| $S_0$, ha | $2.7 \pm 0.7$ | $8.2 \pm 5.6$ | $9.2 \pm 4.5$ | $2.2 \pm 0.9$ | $10.2 \pm 7.5$ | $7.0 \pm 2.4$ | $4.9 \pm 3.4$ |
| $\mu/\sigma^2$ | $0.24 \pm 0.09$ | $0.5 \pm 0.1$ | $0.24 \pm 0.17$ | $1.0 \pm 0.3$ | $0.5 \pm 0.2$ | $0.7 \pm 0.1$ | $0.5 \pm 0.1$ |
| $\lambda/\sigma^2$ | $0.25 \pm 0.06$ | $0.16 \pm 0.08$ | $0.5 \pm 0.2$ | $0.4 \pm 0.2$ | $0.7 \pm 0.2$ | $0.4 \pm 0.1$ | $0.03 \pm 0.02$ |
| $\bar{S}$, ha | 2.0 | 9.0 | 6.9 | 19.9 | 15.0 | 22.9 | 11.5 |
| Median area, ha | 2.0 | 5.9 | 5.5 | 2.9 | 10.0 | 8.7 | 2.8 |
| Mean area, ha | 3.0 | 10.6 | 7.2 | 16.3 | 13.6 | 23.2 | 14.5 |

Table 3. As Table 2 but for various combinations of the adjacent Trypillia stages as indicated in the heading and, in the last two columns, for the whole data set, first for the absolute areas and then, for their normalized values. The median and mean areas from the data are also shown in the last three rows for comparison with the corresponding quantities obtained from the fits.

| Stage | A + BI | A + BI + BI-BII | BII-CI + CI | A–CII [1] | A–CII [2] |
|---|---|---|---|---|---|
| $\alpha$ | 0.7 ± 0.1 | 0.7 ± 0.1 | 0.9 ± 0.1 | 1.0 ± 0.1 | 1.1 ± 0.1 |
| $y_\alpha$ | 3.1 ± 0.2 | 3.7 ± 0.4 | 5.9 ± 0.6 | 6.8 ± 0.3 | 0.3 ± 0.2 |
| $F_\alpha/R^2$ | 0.003/0.97 | 0.04/0.92 | 0.001/0.90 | 0.0002/0.98 | 0.0007/0.99 |
| $\beta$ | 1.7 ± 0.4 | 1.4 ± 0.3 | 1.1 ± 0.1 | 1.2 ± 0.2 | 1.8 ± 0.2 |
| $y_\beta$ | 0.9 ± 0.2 | 1.2 ± 0.3 | 1.6 ± 0.1 | 3.1 ± 0.3 | 16.7 ± 1.9 |
| $F_\beta/R^2$ | <u>0.14</u>/0.95 | 0.05/0.89 | 0.0005/0.96 | 0.004/0.90 | 0.02/0.97 |
| $S_0$, ha | 2.6 ± 0.5 | 3.3 ± 1.1 | 9.1 ± 3.2 | 5.5 ± 1.4 | 0.004 ± 0.03 [3] |
| $\mu/\sigma^2$ | 1.0 ± 0.2 | 0.9 ± 0.2 | 0.6 ± 0.1 | 0.6 ± 0.1 | 0.8 ± 0.1 |
| $\lambda/\sigma^2$ | 0.6 ± 0.1 | 0.5 ± 0.2 | 0.5 ± 0.1 | 0.6 ± 0.1 | 1.0 ± 0.1 |

**Notes:** (1) areas in hectares; (2) areas normalized to the total area of permanent settlements in each stage; (3) dimensionless area normalized to the total area in each individual stage.

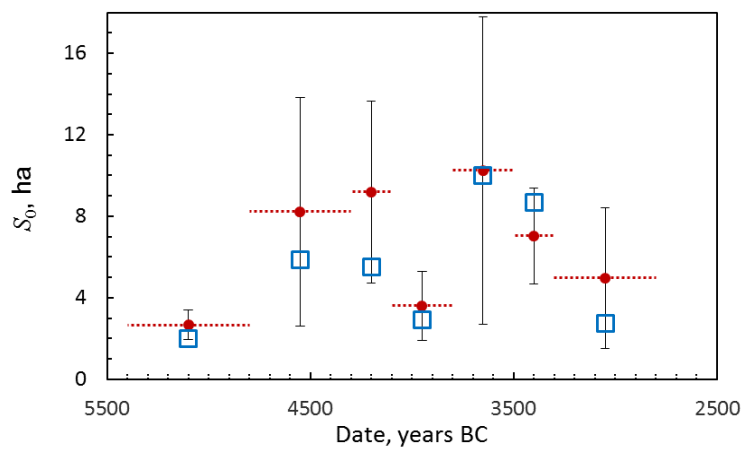

Figure 11. Filled circles with vertical error bars show the initial area of newly formed settlements in each stage, as given in Table 2, and open squares represent the median areas of permanent settlements (computed excluding the sites that have not been included into the power-law fits and this shown with open circles in Figure 8). Horizontal dotted lines indicate the duration of each stage.

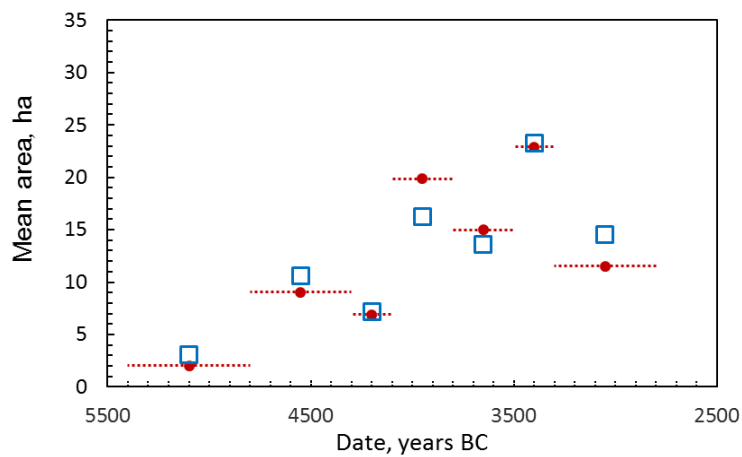

Figure 12. The mean area of permanent settlements in each stage, estimated from the fitted model of stochastic growth as given in Equation (12) and shown in Table 2 (filled circles), and observed (open squares), computed excluding the sites that have not been included into the power-law fits and this shown with open circles in Figure 8.

## 7.1. *The area of newly formed settlements*

The average starting area of a settlement is shown as a function of time in Figure 11. Relatively small values of $S_0 = 2$ and 5 ha are notable in the earliest and latest stages A and CII (and also BII), but $S_0$ are larger at 7–10 ha in the other stages. The measured median area of permanent settlements from Table 1, is also shown in Figure 11 with open squares. The agreement is striking in all stages. Since the values of $S_0$ obtained from the fits and the median areas in the observed area distributions rely on entirely different arguments and calculations, this provides a weighty confirmation of the adequacy and accuracy of the model of stochastic settlement growth supplemented with the emergence of new settlements.

The similarity between the median and initial settlement areas suggests that the newly formed settlements grew or decayed at nearly equal probabilities. In Stages BI, BI-BII and CII, more than half of the new settlements decayed, whereas the opposite trend is noticeable in Trypillia CI. Shukurov et al. (2015) discuss pre-modern farming economy in the Cucuteni–Trypillia area, suggesting that a farming village of 2 ha in area with about 50 inhabitants had its arable fields within 0.5 km of the settlement boundary and the livestock grazing area within 2 km. These distances depend on the human diet structure, fraction of fallow land and other parameters, but the dependence is not very strong for reasonable parameter values. According to the palaeoeconomy model, 2 ha appears to be an optimal size of a village.

The exploitation area of a farming village is limited by the time required to travel to the fields and grazing areas, with cultivated fields within 5 km of a settlement (preferably within 1–2 km) and the livestock kept within 5 km (Chisholm, 1979: 72; Higgs and Vita-Finzi, 1972; Jarman et al., 1982). A village of about 10 ha in area (with assumed number of inhabitants of 270) is expected to have cultivated fields within 1.2 km and grazing areas, within 5 km (Shukurov et al., 2015). This appears to be the maximum plausible size of a newly formed village, in agreement with the estimated values of $S_0$, which do not exceed 10 ha. We also note that a plausible find of an antler ard at the Maidanetske II–Grebenukiv Yar site is dated to Trypillia BI (Pashkevich and Videiko, 2006). The use of ard relieves the seasonal time stresses of agricultural production, especially in the land preparation for sowing. The introduction of the ard may have contributed to the increase of the size of newly formed villages apparent in Figure 11 after Trypillia BI (with the exception of BII).

Figure 12 shows the mean area of the permanent settlements in each stage, both derived from the fitted model and observed. The agreement between the mean areas obtained from the fits and directly from the data confirm that the fits are of sufficiently high quality and represent correctly the overall pattern despite the omission of some data points shown with open circles in Figures 6–9.

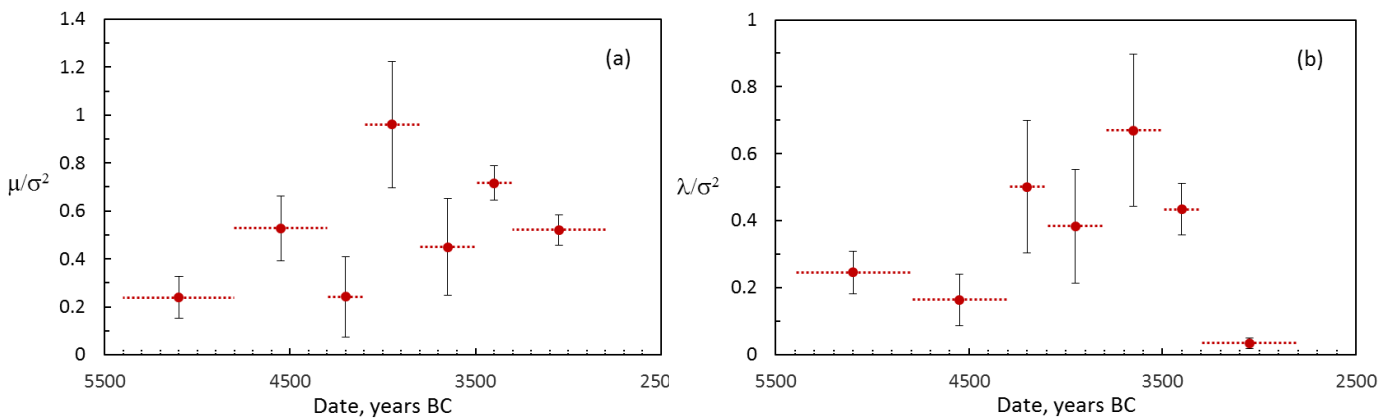

Figure 13. The growth rates of the settlement areas (a) and their number (b), both normalized to their variance. The duration of each stage is shown with horizontal dotted lines.

## 7.2. *Expansion of the settlement system*

The values of $\mu/\sigma^2$ and $\lambda/\sigma^2$ given in the last two rows of Table 2 and shown in Figure 13 attest to a systematic growth in both the areas of individual settlements and their number. There are, however, notable differences in the two processes, visible especially clearly in Figure 13. The relative growth rate of the number of settlements, $\lambda/\sigma^2$, is generally comparable to $\mu/\sigma^2$, especially at the later stages; both are in the range 0.5–0.8. This is not surprising as both growth rates are likely to be related to the rate of the population growth. There are signs of a systematic increase in both parameters during Stages A–CI, but the errors are too large to be certain if this trend is real. With the exception of Trypillia BI-BII and BII-CI where $\mu < \lambda$, the total settlement area grows faster than the number of settlements or at the same rate in Stage A, $\mu \geq \lambda$ (see Figure 14). Thus, a predominant trend was the growth of existing settlements, whereas the formation of new settlements played a less important role as the population grew. The maximum values of both $\mu/\sigma^2$ and $\lambda/\sigma^2$ occur during Stages BI–CI: this was the optimal period where the settlement system was structurally stable with the growth in the number of settlements roughly matching their area growth. The area of the settlements is directly connected to the size of the population but the formation of new settlements is a subtler process. Indeed, another trend, pronounced stronger, is the abrupt decrease, almost to zero, in the rate of growth of the number of settlements during Stage CII. The growth of the settlement area is also slower in CII than in the earlier stages, but the reduction is not that dramatic: the ratio of the two growth rates is as large as $\mu/\lambda = 15 \pm 7$. During that stage, the formation of new settlements was strongly suppressed and the total settlement area grew at a lower rate than earlier, and mainly in pre-existing settlements.

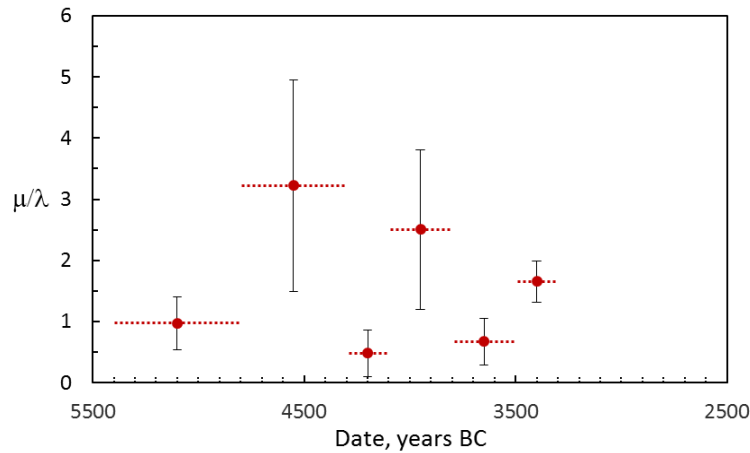

Figure 14. The ratio of the growth rates of the total area of the permanent settlements $\mu$ to their number $\lambda$. The Stage CII data point, $\mu/\lambda = 15 \pm 7$, is much higher than the other ones and is not shown.

The estimated values of $\mu/\sigma^2$ can be used to derive $\mu$ and $\sigma^2$ separately using the growth rate averaged over the stages from A to CI that is estimated in Section 5.1 from the total areas of the settlements, $\bar{\mu} \approx 1/(460 \pm 100 \text{ years})$. Averaging the values of $\mu/\sigma^2$ from Table 2 over the same period, we have $\overline{\mu/\sigma^2} = 0.52 \pm 0.26$ where the range represents the standard deviation around the mean value. Then $\overline{\sigma^2} = \bar{\mu}/\overline{\mu/\sigma^2} = 1/(240 \pm 180 \text{ years})$. Similar calculation using the growth rate of the number of settlements yields a slightly different value, $\overline{\sigma^2} = \bar{\lambda}/\overline{\lambda/\sigma^2} = 1/(410 \pm 320 \text{ years})$, which is, however, consistent with the result obtained from $\mu$ within errors. The average value of the time scale of the variations in the expansion rate of the settlement system is of order $1/\sigma^2 \approx 300$ years, comparable to the duration of a stage. For comparison, Gabaix (1999: 745) quotes the values of $1/\sigma^2$ ranging from about 800 years for the cities in France and Japan in the twentieth century to 200 years for the UK in 1800–1850 and to 50 years for the US cities in the twentieth century.

The rate of growth in the total area of the settlements, $\bar{\mu} \approx 1/460$ years, differs significantly from the likely rate of the population growth, about $\lambda \approx 1/30$ years. If most of the permanent settlements were discov-ered, and their areas were estimated accurately, the difference would suggest an implausibly rapid and strong increase in the population density within the settlements as $\exp[(\gamma - \bar{\mu})t] \approx \exp(\gamma t)$. This suggests that a significant fraction of permanent settlements remains unknown and this fraction is larger in the later stages. Therefore, any estimates of the Trypillian population obtained from the area of known settlements are highly questionable.

## 8. Discussion and conclusions

The agreement between the model of stochastic settlement growth of Reed (2002) and the system of Trypillian settlements, demonstrated in this paper, is striking. Particularly impressive is the agreement between the size of newly formed settlements with the measured mean settlement area in *each* phase of the evolution of the Trypillian Culture. The success of the model strongly suggests that a probabilistic description of the evolution of prehistoric settlement systems provides a viable general framework for the interpretation of settlement hierarchies in terms of social, political, economic processes. The conceptual basis of this approach is the description of numerous, independent and unidentifiable social and demographic processes in terms of the probabilities of the growth, decay and fission of settlements. Then mathematical modelling provides reliable tools to isolate generic features of a settlement system that are independent of the details of the background social and political processes, and then to identify the culturally informative aspects of the data.

Our results strongly suggest that deviations from Zipf's law in prehistoric settlement systems should not be straightforwardly interpreted as evidence of political and administrative centralisation or decentralisation. More plausibly, they are due to the evolution of the settlement system. We have also demonstrated that adequate representation of the data, before any analysis is attempted, can dramatically affect the course and conclusions of such analysis.

The model of stochastic growth provides a quantitative explanation of the Trypillian settlement hierarchy for all stages, and leads to consistent results for the size of newly formed settlements as well as the growth rates of the number and total area of the settlements *in each typo-chronological phase*. The difference between the growth rates of all permanent settlements and those with known area suggests that the recovery rate of the settlements is not the same for all stages. This may have affected the estimates of the mean growth rates of both areas and numbers of the settlements together with other results. Thus, our results should be treated with caution and carefully compared to the archaeological and other evidence. On the other hand, the systematic and transparent nature of our results provide some confidence that our results are generally reasonable.

Trypillia B-BII saw the fastest increase in the total settlement area, whereas the number of settlements increased most rapidly in Stage CI. The terminal Stage CII of the Trypillian Culture is notable for the smallest size of the newly formed settlements (as well as of their mean and median areas) among all stages except for the initial Stage A. The rate of emergence of new settlements is anomalously low in Trypillia CII, ten times lower than in the preceding stages. These are clear signatures of stress. The size distribution of the CII settlements is unusually flat because of the relatively large number of small and moderate-size sites, unlike the other stages where the frequency of site areas has a pronounced maximum at 2–10 ha. The largest known site of Trypillia CII is 160 ha in area, as opposed to 341 ha in the previous stage CI. It appears that the population was dispersed into smaller settlements in CII.

The evolution in the quantitative parameters of the Trypillian settlement system undoubtedly can be interpreted in terms of increased internal stress and conflict in larger settlements which might promote their fission and creation of smaller sites. On the other hand, external stress and hostilities are likely to facilitate nucleation in larger, defended settlements and their location away from main rivers (see Duffy, 2015, and references therein). However, any assessment of the social and political implication of the formal data analysis should to be performed with full allowance for the archaeological, palaeoenvironmenal and other evidence. Such an assessment is beyond the scope of this paper. Inter-pretations of the settlement sizes in terms of quantitative models, similar to that performed here, provides a solid basis for such analyses. In this sense, mathematical modelling provides a unique opportunity for reliable representation of the data and identification of its informative features.

## Acknowledgements

We are grateful to Dr Kavita Gangal and Dr Andrew Baggaley for useful discussions and for preparing the maps shown in Figure 2. Helpful comments of G. R Sarson are gratefully acknowledged. AS acknowledges financial support of the Leverhulme Trust (grant RPG-2014-427) during this research and writing the paper. MV is grateful to the School of Mathematic and Statistics of Newcastle University (Prof. R. Henderson) for financial support and hospitality.

**Conflicts of Interest:** The authors declare that they have no conflict of interest.

## References

Batty, M., 2006. Rank clocks. *Nature*, 444, 592–596.

Berry, B. J. L., and Okulicz-Kozaryn, A., 2012. The city distribution debate: resolution for US urban regions and megalopolitan areas. *Cities*, 29, S17–S23.

Brackman, S., Garretsen, H., Van Marrewijk, C., and van den Berg, M., 1999. The return to Zipf: towards a further understanding of the rank–size distribution. *Journal of Regional Science*, 39, 183–213.

Brigand, R., and Weller, O., 2013. Neolithic and Chalcolithic settlement patterns in central Moldavia (Romania). *Documenta Praehistorica*, XL, 195–207.

Burdo, N. B., 2003. Sakralnyi aspekt arhitektury tripolskikh protogorodov. In: Korvin-Piotrovsky, A. G. (Ed.), *Tripilski Poselennya-Giganty. Materialy Mezhdunarodnoi Konferencii,* Kyiv, Korvin-Press, pp. 8–13 (in Russian).

Burdo, N. B., 2006. Sakralnyi svit i magicnyi prostir trypilskoi civilizacij. In: Zubar, V. M. (Ed.) *Arkheologia u Kievo-Mogylanskiej Akademij*, Kyiv, Stilos, pp. 75–87 (in Ukrainian).

Burdo, N. B., Videiko, M. V., 2005. Zhivotnovodstvo tripol'koj kul'tury v period 5400–2750 gg. do n.e. (Livestock breeding in the Trypillian Culture of 5400–2750 BC). In Kośko, A., Szmyt M. (Eds.), *Nomadysm a pastoralism w międzyrzecu Wisly i Dniepru (neolit, eneolit, epoka brązu)*, Arheologia Bimaris, Poznań, Wydawnictwo Poznańskie, Vol. 3, pp. 67–93.

Burdo, N., Videiko, M., 2016. Nebelivka: from magnetic prospection to new features of mega-sites. In: Müller, J., Videiko, M., Rassmann, K. (Eds.) *Trypillia – Megasites and European Prehistory*, London and New York: Routledge, pp. 95–116

Burdo N., Videiko M., Chapman J., and Gaydarska B., 2013. Houses in the archaeology of the Trypillia–Cucuteni groups. In Hofmann, D., Smyth, J. (Eds.) *Tracking the Neolithic House in Europe. One World Archaeology*, New York and London: Springer Science, pp. 95–116

Chapman, J., Videiko, M., Hale, D., Gaydarska, B., Burdo, N., Rassmann, K., Mischka, C., Müller, J., Korvin-Piotrovskiy, A., and Kruts, V., 2014. The second phase of the Trypillia mega-site methodological revolution: a new research agenda. *European Journal of Archaeology*, 17, 369–406.

Chisholm, M., 1979. *Rural Settlement and Land Use: an Essay in Location* (3rd ed.). London, Hutchinson.

Dergachev, V.A., 1980. *Pamiatniki Pozdnego Tripolia* (Monuments of the Late Tripolye). Kishinev, Shtiinţa (in Russian).

Dewar, R. E., 1991. Incorporating variation in occupation span into settlement-pattern analysis. *American Antiquity,* 56, 604–620.

Dewar, R. E., 1994. Contenting with contemporaneity: a reply to Kintigh. *American Antiquity,* 59, 149–152.

Diachenko, O. V., 2008. Dynamika zmin chisel'nosti naselenija volodimirivs'ko–tomashivs'kij linii zakhidnotripil'kij kul'turi [The dynamics of population change of the Volodimirovka–Tomashivka branch of the western Trypillia culture]. *Arkheologija*, No. 4, 9–17 (in Ukrainian).

Diachenko, O. V., 2009. Prostorova organizacia zakhidnotrypilskogo suspilstva u mezhyricchi Pivdennogo Bugu i Dnipra. *Arkheologichni Stidii, Magisterium,* 36, 19–26 (in Ukrainian).

Diachenko, O. V., 2010. *Trypilske naselennya Bugo-Dniprovskogo mezhyricchia: proostorovo-chasovyj analiz. PhD Thesis*, Kyiv (in Ukrainian).

Drennan, R. D., and Peterson, C. E., 2004. Comparing archaeological settlement systems with rank-size graphs: a measure of shape and statistical confidence. *Journal of Archaeological Science*, 31, 533–549.

Dudkin, V.P., and Videiko, M. Yu., 2009. *Arkhitektura Trypilskoi Tsivilizacij: vid Poselen do Proto-Mist,* Kyiv, Myslene Drevo (in Ukrainian).

Duffy, P. R., 2014. *Complexity and Autonomy in Bronze Age Europe: Assessing Cultural Development in Eastern Hungary.* Archaeolingua, Budapest.

Duffy, P. R., 2015. Site size hierarchy in middle-range societies. *Journal of Anthropological Archaeology,* 37, 85–99.

Eeckhout, J., 2004. Gibrat's law for (all) cities. *American Economic Review*, 94, 1429–1451.

Flannery, K. V., 1976. Evolution of complex settlement systems. In Flannery, K. V. (Ed.), *The Early Mesoamerican Village.* Academic Press, New York, pp. 162–173.

Gabaix, X., 1999. Zipf's law for cities: an explanation. *Quarterly Journal of Economics*, 114, 739–767.

Gabaix, X., 2009. Power laws in economics and finance. *Annual Review of Economics*, 1, 255–293.

Gabaix, X., and Ioannides, Y. M., 2004. The evolution of city size distributions. In Henderson, J. V., Thisse, J.-F. (Eds.), *Handbook of Regional and Urban Economics, Vol. 4: Cities and Geography.* Elsevier Science, Oxford, pp. 2341–2378.

Gardiner, C., 2009. *Stochastic Methods. A Handbook for Natural and Social Sciences*. Berlin, Springer.

Giesen, K., Zimmermann, A., and Suedekum, J., 2010. The size distribution across cities – double Pareto lognormal strikes. *Journal of Urban Economics*, 68, 129–137.

Griffin, A. F., 2011. Emergence of fusion/fission cycling and self-organized criticality from a simulation model of early complex polities. *Journal of Archaeological Science*, 38, 873–883.

Harper, T. K., 2012. K problemie razmerov poselenia u s. Talianki [On the problem of the settlement size near the Talianki village]. http://www.acsu.buffalo.edu/~tkharper/Harper2012rus.pdf.

Higgs, E. S., and Vita-Finzi, C., 1972. Prehistoric economies: a territorial approach. In Higgs, E. S. (Ed.), *Papers in Economic Prehistory*. Cambridge, Cambridge University Press, pp. 37–36.

Jarman, M. R., Bailey, G. N., and Jarman, H. N. (Eds.), 1982. *Early European Agriculture. Its Foundations and Development*. Cambridge, Cambridge University Press.

Johnson, G. A., 1977. Aspects of regional analysis in archaeology. *Annual Review of Anthropology*, 6, 479–508.

Johnson, G. A., 1980. Rank-size convexity and system integration: a view from archaeology. *Economic Geography*, 56, 234–247.

Khvoika, V., 1904. Raskopki 1901 g. v oblasti tripolskoj kultury [Excavations of 1901 in the Trypolye Culture area]. *Zapiski Otdelenia Russkoj i Slavianskoj Arkheologii Imperatorskogo Russkogo Arkheologicheskogo Obschestva,* 5, 12–20.

Kintigh, K. W., 1994. Contending with contemporaneity in settlement-pattern studies. *American Antiquity,* 59, 143–148.

Knappett, C., Evans, T., Rivers, R., 2008. Modelling maritime interaction in the Aegean Bronze Age. *Antiquity*, 82, 1009–1024.

Kohler, T. A., and Varien, M. D., 2010. A scale model of seven hundred years of farming settlements in Southwestern Colorado. In Bandy, M. S., Fox, J. R. (Eds.), *Becoming Villagers. Comparing Early Village Societies*. Univ. Arizona Press, Tucson, pp. 37–61.

Krutz, V. A., 1989. K istorii naselenija tripol'skoy kultury v mezhdurech'e Yuzhnogo Buga i Dnepra [On the history of the Tripolyian population in the Dnieper–Southern Bug interfluves]. In *Pervobytnaya Arkheologiya. Matrialy i Issledivanija [Prehistoric Archaeology. Materials and Studies]*. Kiev, Naukova Dumka, pp. 117–132 (in Russian).

Malevergne, Y., Pisarenko, V., and Sornette, D., 2011. Testing the Pareto against the lognormal distributions with the uniformly most powerful unbiased test applied to the distribution of cities. *Physical Review E*, 83, 036111.

Malevergne, Y., Saichev, A., and Sornette, D., 2013. Zipf's law and maximum sustainable growth. *Journal of Economic Dynamics and Control*, 37, 1195–1212.

Markevich, V. I., 1981. *Pozdnetripolskiye plemena severnoy Moldavii* [The Late-Tripolyian Tribes of Northern Moldavia]. Kishinev, Stiinitsa (in Russian).

Müller, J., Hofmann, R., Brandstatter, L., Ohlrau, R., and Videiko, M., 2016. Chronology and demography: how many people lived in a mega-site? In Müller, J., Videiko, M., Rassmann, K. (Eds.), *Trypillia – Megasites and European Prehistory*, London and New York, Routledge, pp. 133–169

Müller, J., and Videiko, M., 2016. Maidanetske: new facts of a mega-site. In Müller, J., Videiko, M., Rassmann, K. (Eds.), *Trypillia – Megasites and European Prehistory*, London and New York, Routledge, pp. 71–93

Nam, K.-M., and Reilly, J. M., 2013. City size distribution as a function of socioeconomic conditions: An eclectic approach to downscaling population. *Urban Studies,* 50, 208–225.

Ortman, S. G., Cabaniss, A. H. F., Sturm, J. O., and Battencourt, L. M. A., 2014. The Pre-History of Urban Scaling. *PLOS One*, 9 (2), e87902 (10 pp.)

Ortman, S. G., Cabaniss, A. H. F., Sturm, J. O., and Battencourt, L. M. A., 2015. Settlement scaling and increasing returns in an ancient society. *Science Advances*, 1: e1400066 (8 pp.)

Pashkevich, G. A., and M. Yu. Videiko, 2006. *Ril'nictvo Plemen Trypil'skoi Cul'tury* [Agriculture of the Tribes of the Trypillia Culture]. Kiev, Naukova Dumka (in Ukrainian).

Passek, T., 1935. La ceramique tripolienne. *Bulletins de l'Académie d'Histoire de la Culture Matérielle,* 122. Moscou: Editions d'etat, Section sociale et economique.

Passek, T. S., 1949. Periodizaciya tripolskikh poselenij (III-II tysiacheletia do n.e.) [Periodisation of Trypillian settlments (III–II millennia BC)]. *Materialy i Issledovania v Arkheologij SSSR,* 10. Moskva-Leningrad (in Russian).

Passek, T. S., and Krychevskij, E. Yu., 1946. Tripolskoe poselenie Kolomyjschyna: opyt rekonstrukcij [The Trypillian settlement Kolomyjschyna : reconstruction attempt]. *Kratkie Soobschenia Instituta Istorii Marerialnoj Kultury*, 12, 14–22 (in Russian).

Pumain, D., Guerois, M., 2004. Scaling laws in urban systems. *Working Papers, Santa Fe Institute*, Santa Fe, MN, p. 4.

Rassmann, K., Korvin-Piotrovskiy, A., Videiko, M., Müller, J., 2016. The new challenge for site plans and geophysics: revealing the settlement structure of giant settlements by means of geomagnetic survey. In: Müller, J., Videiko, M., Rassmann, K. (Eds.), *Trypillia-Megasites and European Prehistory*, London and New York, Routledge, pp. 29–54.

Reed, W. J., 2001. The Pareto, Zip and other power laws. *Economics Letters*, 74, 15–19.

Reed, W. J., 2002. On the rank–size distribution for human settlements. *Journal of Regional Science*, 42, 1–17.

Reed, W. J., 2003. The Pareto law of incomes – an explanation and an extension. *Physica A*, 319, 469–486.

Reed, W. J., and Hughes, B. D., 2002a. On the size distribution of live genera. *Journal of Theoretical Biology*, 217, 125–135.

Reed, W. J., and Hughes, B. D., 2002b. From gene families and genera to incomes and internet file sizes: Why power laws are so common in nature. *Physical Review E*, 66, 067103 (4 pp.).

Reed, W. J., and Jorgensen, M., 2004. The double Pareto-lognormal distribution – a new parametric model for size distributions. *Communications in Statistics – Theory and Methods*, 33, 1733–1753.

Rozenfeld, H. D., Rybski, D., Gabaix, X., and Makse, H. A., The area and population of cities: new insights from a different perspective on cities. *American Economic Review*, 101, 2205–2225.

Savage, S. H., 1997. Assessing departures from log-normality in the rank–size rule. *Journal of Archaeological Science,* 24, 233–244.

Shishkin, K. V., 1973. O praktuke deshyfrovaniya aerofotoznimkov v arkheologichykh tselyakh [On the practice of interpretation of aerial photographs for archaeology]. *Arkheologia,* 10, 32–41.

Shishkin, K. V., 1985. Planyvannya trypilskikh poselen za danymi aerofotoziomki [Plans of Trypillian settlements according to aerial photography data]. *Arheologia,* 52, 72–77 (in Ukrainian).

Shmaglij, M. M., and Videiko, M. Yu., 1993. Trypilski protomista [Trypillian proto-cities], *Arkheologia,* 3, 52–63 (in Ukrainian).

Shmaglij, M. M., Dudkin, V. P., and Zinkovskij K. V., 1973. Pro kompleksne vyvchennya trypilskikh poselen [On complex studies of Trypillian settlements], *Arkheologia,* 10, 23–31 (in Ukrainian).

Shukurov, A., Videiko, M. Y., Sarson, G. R., Davison, K., Shiel, R., Dolukhanov, P. M., and Pashkevich, G. A., 2015. Productivity of pre-modern agriculture in the Cucuteni–Trypillia area. *Human Biology*, 87, Article 8, 235–282.

Siegel, S., and Castellan, N. J., 1988. *Nonparametric Statistics for the Behavioral Sciences*. McGraw-Hill, New York.

Skakun, N. N., 2005. Bodaky – odin iz tsentrov kremneobrabatyvajuschego proizvodstva na Volyni [Bodaki, one of flint processing centres in Volyn']. In Skakun N. N., Krutz V. A., Mateva B., Korvin-Piotrovski A. G., Samzun A., Yakovleva L. M. (Eds.), *Arkheologicheskie issledovania tripolskogo poselenia Bodaki v 2005 godu*. Korvin-Press, Kiev–Sankt-Petersburg, pp. 67–79 (in Russian).

Sumner, W. M., 1989. Population and settlement area: An example from Iran. *American Anthropologist*, 91, 631–641.

Telegin, D. Ya., 1985. Radiokarbonne i arkheomagnitne datuvann'a tripil'skoy kul'turi [The radiocarbon and archaeomagnetic dating of the Trypillian culture]. *Arkheologiya: Respublik. Mezhvid. Zb. Nauk Pr. [Archaeology: National Interdisciplinary Collection of Research Papers]*, No. 52, Kyiv, pp. 10–22 (in Ukrainian).

Tsvek, E. V., 2006. *Poselennya Shidnotrypilskoi kultury (korotkyi narys) [Settlements of the Western Trypillian Culture: A Brief Sketch]*. Kyiv, KP OTI (in Ukrainian).

Vasylenko, B. A., 1989. Vydobyvannya i obrobka kremeniy na pravoberezhzhi Verkhniego Podnistrovia v eneoliti. *Problemy Istorii ta Arkheologii Davnyago Naselennya Ukrainskoi RSR, tez. dop. XX Resp. Konf.*, Kyiv, pp. 38–39 (in Ukrainian).

Videiko, M. Yu., 1992. *Ekonomika ta Syspilnyj lad Trypilskogo Naselennya Pivdennogo Pobuzhzhya (Etapy BII-CI). PhD Theis*, Kyiv (in Ukrainian).

Videiko, M. Yu., 2005. Trypilske poselenie Ignatenkova Gora bilya s. Grygorivka. *Kamiana Doba Ukrainy*, 7, 186–200 (in Ukrainian).

Videiko, M. Yu., 2013. *Kompleksnoie Izuchenie Krypnykh Poselenij Tripolskoj Kultury. V–IV Tysiacheletia do n.e.* [Complex Exploration of the Large Settlements of the Trypillian Culture or the V–IV Centuries BC], Saarbruken, Lambert Publ. (in Russian).

Videiko, M. Yu., 2015. *Pivdenno-Skhidna ta Tsentralna Evopa u V–IV Tys. do n.e.* [South-Western and Central Europe in V–IV Centuries BC]*,* Kyiv–Uman (in Ukrainian).

Videiko, M. Yu., 2016. Dnestro-Dneprovskoye mezhdurechie na rubezhe V i IV tysiacheletij do n.e., *Stratum Plus*, 2, 64–67 (in Russian).

Videiko, M. Yu., N. B. Burdo and S. M. Lyashko (eds), 2004. *Entsiklopedia Tripil'skoy Tsivilizatsii* [*An Encyclopaedia of the Trypillian Civilization*], Vols 1 and 2. Kyiv, Ukrpoligrafmedia (in Ukrainian).

Videiko, M., Chapman, J., Burdo, N., Geydarska B., Ignatova S., Ivanova S., and Rud V., 2014. Issledovania po proektu "Rannia urbanizacia v praistoricheskoj Evrope? Tripolskie megaposelenia" v 2013 godu, *Tyragetia,* VIII [XXIII], 1, 107–144 (in Russian).

Wilshusen, R. H., and Potter, J. M., 2010. The emergence of early villages in the American Southwest. In Bandy, M. S. , Fox, J. R. (Eds.), *Becoming Villagers. Comparing Early Village Societies.* Univ. Arizona Press, Tucson, pp. 165–183.

Zbenovich, V. G., 1980. *Poselenie Bernashevka na Dnestre: k Proiskhozhdeniu Tripolskoj Kultury.* Kyiv, Naukova Dumka (in Russian).

Zbenovich, V. G., 1989. *Rannii Etap Tripolskoj Kultury na Territorii Ukrainy.* Kyiv, Naukova Dumka (in Russian).